\documentclass[modern]{aastex63}

\renewcommand{\deg}{^{\circ}}
\received{01/23/2020}
\revised{03/16/2020}
\accepted{}
\submitjournal{AJ}

\shorttitle{Radial distribution of magnetic flux ropes}
\shortauthors{Chen and Hu} 

\graphicspath{{FIGS/}}
\begin{document}

\title{Effects of Radial Distances on
Small-scale Magnetic Flux Ropes in the Solar Wind}

\correspondingauthor{Qiang Hu}
\email{qiang.hu@uah.edu, qiang.hu.th@dartmouth.edu}


\author[0000-0002-0065-7622]{Yu Chen}
\affiliation{Department of Space Science \\
The University of Alabama in Huntsville \\
Huntsville, AL 35805, USA}

\author[0000-0002-7570-2301]{Qiang Hu}
\affiliation{Department of Space Science \\
Center for Space Plasma and Aeronomic Research (CSPAR) \\
The University of Alabama in Huntsville \\
Huntsville, AL 35805, USA}

%
%
%
%
%
%
%



\begin{abstract}
Small-scale magnetic flux ropes (SFRs), in the solar wind,  have
been studied for decades. Statistical analysis utilizing various
in situ spacecraft measurements is the main observational approach
which helps investigate the generation and evolution of these
small-scale structures.  Based on the Grad-Shafranov (GS)
reconstruction technique, we use the automated detection algorithm
to build the databases of these small-scale structures via various
spacecraft measurements at different heliocentric distances. We
present the SFR properties including the magnetic field and plasma
parameters at different radial distances from the sun near the
ecliptic plane. It is found that the event occurrence rate is
still in the order of a few hundreds per month, the duration and
scale size distributions follow power laws, and the flux rope axis
orientations are approximately centered around the local Parker
spiral directions. In general, most SFR properties exhibit radial
decays. In addition, with various databases established, we derive
scaling laws for the changes of average field magnitude, event
counts, and SFR scale sizes, with respect to the radial distances,
ranging from $\sim$ 0.3 au for Helios to $\sim$ 7 au for the
Voyager spacecraft. The implications of our results for
comparisons with the relevant theoretical works and for the
application to the Parker Solar Probe (PSP) mission are discussed.

\end{abstract}




\section{Introduction} \label{sec:intro}
In the past few decades, there has been a number of observational
studies on small-scale magnetic flux ropes (hereafter, SFRs) in
the solar wind. These relatively small-scale structures, first
identified from  the \emph{Ulysses} spacecraft measurements at
$\sim$ 5 au by \citet{Moldwin1995}, generally are believed to have
the same helical magnetic field configuration as their large-scale
counterparts, i.e., magnetic clouds (MCs). Employing various
in-situ spacecraft mission datasets, such as those from
\emph{IMP-8}, \emph{WIND} and \emph{STEREO} at about 1 au, a
limited number (usually in the order of hundreds in total at most)
of SFRs were identified for multiple  years via mostly visual
inspection or semi-automated method. For example,
\citet{Cartwright2010} used \emph{Helios 1 $\&$ 2}, \emph{IMP-8},
\emph{WIND}, \emph{ACE} and \emph{Ulysses} spacecraft datasets
with the corresponding heliocentric distances ranging from
$\sim$0.3 to 5.5 au to investigate the occurrence and the
evolution of hundreds of SFRs. This represents the first
comprehensive study of SFRs and the generation of  SFR databases
from multiple spacecraft missions, although the event counts are
very limited.  The corresponding statistical studies on these
limited-size samples suggested that these small structures can
last for about tens of minutes up to a few hours, and thus have
smaller scale sizes over their cross sections as compared with
magnetic clouds \citep{Moldwin2000}.

 In the
meanwhile, the distinctions between the two populations were
sought out, and the two possible sources, i.e., small coronal mass
ejection (CME) from solar eruption and magnetic reconnection
across the heliospheric current sheet (HCS) in the solar wind,
were proposed to be origins of SFRs
\citep{Feng2007,Feng2008,CartWright2008}. In addition,
\citet{Yu2016,Yu2014} performed careful and detailed analysis of
in-situ observations of small transients (STs) including SFRs in
the solar wind around 1 au. They examined their overall magnetic
and plasma properties, and  concluded that these relatively
small-scale structures may originate from both the solar corona
and the interplanetary medium.

One step forward in the study of SFRs is the implementation of
automated and computerized algorithm, based on the Grad-Shafranov
(GS) reconstruction technique, for identifying these structures
from in-situ spacecraft measurements. This has yielded
 significantly more number of events, which is more substantial for statistical
investigations. The GS method  \citep[see][for a comprehensive
review]{Hu2017GSreview} is a unique data analysis technique
capable of recovering the two dimensional (2D) structure from one
dimensional (1D) time series data
\citep{Sonnerup1996,Hau1999,Hu2001,Hu2002}. \citet{zhengandhu2018}
created the  computer-based program to identify SFRs
automatically. This automated detection has succeeded in finding
74,241 SFRs by using the \emph{WIND} spacecraft in situ
measurements from 1996 to 2016 \citep{2018ApJS..239...12H}. By
this abundant event count number (in the order of a few hundreds
per month on average), they compared the monthly counts of SFRs
with the corresponding sunspot numbers and indicated that the
occurrence of SFRs has  obvious solar cycle dependency with a
short lag. Later, this automated detection was applied to the
\emph{ACE} and \emph{Ulysses} measurements \citep{Chen2019}. The
bulk properties of identified SFRs, including the magnetic field
strength and plasma parameters, are presented in terms of their
variations with time, heliographic latitudes and radial distances.
It is found that the solar cycle dependency or the temporal
variation of SFRs appears to be affected by both latitudinal and
radial-distance changes owing to the unique orbit of
\emph{Ulysses}. An earlier study on par with the number of events
identified by the GS-based approach was performed by
\citet{Borovsky2008} using \emph{ACE} measurements to identify
boundaries of about 65,860 flux tubes for seven years worth of
data. That study suggested that these flux tubes are tangled along
the Parker spiral direction and forming a scenario made up of
``spaghetti" like structures originating from the Sun.

On the other hand, the other relevant studies have hinted at the
local generation of these flux tube/rope structures from 2D
magnetohydrodynamic (MHD) turbulence. These 2D turbulence is
characterized by quasi-2D coherent structures manifested as
current sheets and flux ropes of variable sizes, corresponding to
spatial  scales in the inertia range
\citep{2007ApJ...667..956M,2008PhRvL.100i5005S,2013PhPl...20d2307W,Zank2017}.
Various studies using in-situ spacecraft measurements
\citep[e.g.,][]{Greco2008,Greco2009a,Osman2014,2018SSRv..214....1G}
were carried out for identifying and characterizing
discontinuities, i.e., current sheets, which may be considered as
proxies to boundaries (or so-called ``walls") of flux ropes. In particular, the partial variance of increments (PVI) method is commonly used to identify discontinuities from in situ magnetic field measurements (see the review by \cite{2018SSRv..214....1G}).
 For instance, the correspondence between the distributions of the wall-to-wall time of SFRs and the waiting time of
  current sheets yielded consistent results in determining the
  correlation length scale for 2D MHD turbulence \citep{zhengandhu2018,Greco2009a}.
Recently, \citet{Pecora2019} examined \emph{WIND} in situ data and
related the SFRs and current sheets by combing the GS
reconstruction and the PVI
methods. They showed the correspondence between the flux rope
boundaries and current sheets, where each type of structures was
identified from the same dataset but with different and
independent approaches.   They certified again that these
small-scale structures can be generated self-consistently from
quasi-2D MHD turbulence.


To further extend and complete our SFR event databases for the
existing and past spacecraft missions, we apply our automated
detection algorithm to \emph{Helios 1 $\&$ 2} and \emph{Voyager 1
$\&$ 2} datasets in the present study. Considering that they offer
observations at additional heliocentric distances complementary to
the \emph{ACE} and \emph{Ulysses} missions, we will also perform a
comprehensive analysis of the possible evolution of SFR properties
with radial distances between $\sim$ 0.3 and 8 au, especially for
a uniquely controlled subset of events to be described below.

This paper is organized as follows. The method of the automated
detection will be described briefly together with the data
selection for this study  in Section \ref{sec:method}. In Section
\ref{sec:results}, the SFRs identified from the full \emph{Helios}
mission is categorized into three groups by their corresponding
heliocentric distances. One specific year is selected for
identification via the \emph{Voyager} mission. The SFRs properties
including the axis orientation angles, the magnetic field and
plasma parameters, the duration and scale size will be discussed
for each event set, respectively. In Section \ref{sec:evolution},
the radial distributions of SFRs via these two missions as well as
\emph{ACE} and \emph{Ulysses} are presented. The radial effects
associated with the possible scenario of flux rope merging are
discussed. The findings and additional discussions, in particular,
regarding the applications to the \emph{Parker Solar Probe (PSP)}
mission, are summarized in the last section.

\section{Method and Data Selection} \label{sec:method} The data analysis method we employ is the
recently developed automated flux-rope detection algorithm based
on the Grad-Shafranov (GS) reconstruction technique. The GS
reconstruction method is an advanced data analysis tool based on
the  two-dimensional (2D) GS equation describing space plasma
structures in approximate 2D quasi-static equilibrium and
employing in-situ spacecraft measurements. It has been widely
applied to various space plasma regimes by a number of research
groups worldwide for over twenty years \citep[see][for a
comprehensive review]{Hu2017GSreview}. The latest development has
been the application of the basic GS reconstruction procedures to
identifying relatively small-scale magnetic flux ropes (SFRs) in
the solar wind in a completely computerized and automated manner
\citep{zhengandhu2018,2018ApJS..239...12H,Chen2019}. This has
enabled the generation of exhaustive event lists for the flagship
NASA solar-terrestrial  spacecraft missions. The event occurrence
rate is in the order of a few hundreds per month, which had not
been achieved by other means before. The detailed documentation of
the approach including an algorithm flowchart was provided in
\citet{2018ApJS..239...12H}, which should enable the
implementation of the algorithm  by interested users on their own.
We provide below a brief description of the basic concepts
underlying the automated detection algorithm, which is also
utilized in the current study.

The basic quantity characterizing the 2D flux rope configuration
is defined through a magnetic flux function, $A(x,y)$, which fully
characterizes the transverse magnetic field on the cross-section
plane $(x,y)$, perpendicular to the flux rope axis, $z$. In other
words, the isosurfaces of $A$, i.e., where $A= const$, represent
flux surfaces on which the magnetic field lines are winding along
the central axis $z$ with $B_z\ne0$ and $\partial/\partial
z\approx 0$. Therefore a solution of the scalar function $A$,
governed by the GS equation, as well as a non-vanishing axial
field component $B_z(A)$ fully characterizes a cylindrical flux
rope configuration with nested flux surfaces surrounding one
central $z$ axis. Thus one important property associated with such
a configuration is the single-valued behavior of the so-called
field-line invariants, i.e., a few quantities as single-variable
functions of $A$ only, i.e., also being constant on each flux
surface. They include the axial field component $B_z$, the plasma
pressure $p$, and the transverse pressure $P_t=p+B_z^2/2\mu_0$.
These quantities together with the $A$ values all can be directly
evaluated along a single spacecraft path once a $z$ axis is
chosen. For a cylindrical flux rope configuration and a spacecraft
path across multiple nested flux surfaces, these quantities, as
single-variable functions of $A$, all exhibit a discernable
double-folding behavior when displayed against $A$ values along
the path. This is because along the spacecraft path, each flux
surface is crossed twice by the spacecraft, once along the inbound
(``1st" half of the) path, while the other outbound (or the ``2nd"
half). Therefore the corresponding pairs of $A$ values are the
same because each is on the same flux surface. So is each pair of
the corresponding invariant quantity, as single-variable function
of $A$,  thus leading to the behavior such that the ``2nd" half of
the data points folds and overlaps with the ``1st" half, becoming
so-called double-folding.

The $A$ values along the spacecraft path (at $y=0$) generally
exhibit a monotonically increasing or decreasing pattern along the
``1st" half, then the trend reverses for the ``2nd" half, after an
extremum is reached. These values $A(x,0)$ are calculated from the
``rotating" component of the magnetic field via $A(x,0)=-\int_0^x
B_y d\xi$, where the spatial increment $d\xi=-\mathbf{V}_F\cdot
\hat{\mathbf x}dt$ is related to a frame velocity $\mathbf{V}_F$
(commonly the average solar wind velocity) and the time increment
$dt$ of the time-series data. The point at which the value
$A(x,0)$ reaches an extremum is called the turning point. It is
also where the component $B_y$ changes sign, separating the ``1st"
and the ``2nd" halves. Therefore the resulting double-folding
behavior in the field-line invariants, as dictated by the GS
equation, constitutes the key feature we utilize to devise the
algorithm for detecting magnetic flux rope intervals from in-situ
spacecraft measurements. The quality of the ``double-folding"
pattern is assessed quantitatively by several metrics to result in
the identification of flux rope candidates. These metrics include
the definition of two residues evaluating primarily the goodness
of the satisfaction for these quantities, in particular the
transverse pressure $P_t$, being single-valued and double-folded.
A Wal\'en slope threshold is also used to exclude mostly
Alfv\'enic fluctuations. In addition, an optional threshold
condition on the average magnetic field magnitude over a candidate
event interval can also be applied to reduce contamination of
small-amplitude fluctuations whose flux-rope characteristics are
less certain. The detailed descriptions of the procedures were
provided in \citet{2018ApJS..239...12H}. We refer interested
readers to that report for further details.

\begin{table}[h]
    \caption{Spacecraft Missions for the Small-scale Magnetic Flux Rope Detection and Analysis.}\label{tbl:missions}
    \centering
    \begin{tabular}{cccccc}
        \hline
        Spacecraft & Helios 1 & Helios 2 & ACE & Ulysses & Voyager 1 (2)\\
\hline
        Periods (years) & 1975-1984 & 1976-1980 & 1998-2018 & 1991-2009 & 1980\\
\hline
    Counts & 15,041 & 7,981 & 38,505 & 22,719 & 1,480 (1,991)\\
    \hline
    \end{tabular}
\end{table}

The automated detection algorithm has been applied to a number of
in-situ spacecraft missions,  including ACE, Wind, and Ulysses
\citep{2018ApJS..239...12H,Chen2019}. A designated website
containing the event databases is available at
\url{http://fluxrope.info}. Due to the implementation of the
highly computerized algorithm and the usage of cluster machines,
the analysis of whole mission dataset becomes feasible. We
continue our analysis in this study for additional in-situ
spacecraft datasets in the heliosphere, specifically the Helios
and Voyager missions. The specific time periods and the resulting
event counts for each mission are listed in
Table~\ref{tbl:missions}. Since we focus on studying the flux rope
properties at different heliocentric distances and attempt to
inter-relate these properties considered to be radially
distributed in the solar wind, we select as many as possible the
periods when data are available for the whole Helios mission, but
only include year 1980 for the Voyagers when they were at large
radial distances, but still in low helio-latitudes near the
ecliptic. Such selections are also largely affected by the data
integrity issues, i.e., the existence of data gaps. When they are
prevalent, especially for plasma parameters which are required in
our analysis, and for relatively historical missions, the data
integrity is significantly reduced and negatively impacts the
search results greatly. Nonetheless, on average, the event
occurrence rate is still in the order of a few thousands  a year,
owing to the fairly exhaustive detection approach, by using
1-minute cadence data throughout \citep{2018ApJS..239...12H}.


\section{Analysis Results} \label{sec:results}
The analysis on a number of heliospheric missions as listed in
Table~\ref{tbl:missions} was carried out systematically for all
available data, except for the Voyager mission. Due to data
integrity issues both concerning the low-quality or missing data
and insufficient resolution of the time series, the analysis
results for Voyagers are limited to year 1980 only. At that time,
the spacecraft were at a heliocentric distance $\sim$ 6.05-9.57 au
near the ecliptic plane. The event counts from the Helios mission
are also relatively lower mostly due to significant number of data
gaps. In Section~\ref{sec:all4}, we present the properties of
identified SFRs for Helios and Voyager missions, respectively. The
detailed descriptions of Ulysses, ACE and Wind results were
already reported elsewhere in \citet{Chen2019} and
\citet{2018ApJS..239...12H}. In Section~\ref{sec:oneyear}, we
present the results from a subset of events for all four missions
under similar condition of over exact one-year time period near
the ecliptic plane, but at different heliocentric distances.

\subsection{Distributions of Selected Flux Rope Properties at Different Radial
Distances}\label{sec:all4}

\begin{figure}
\centering
\includegraphics[width=.48\textwidth]{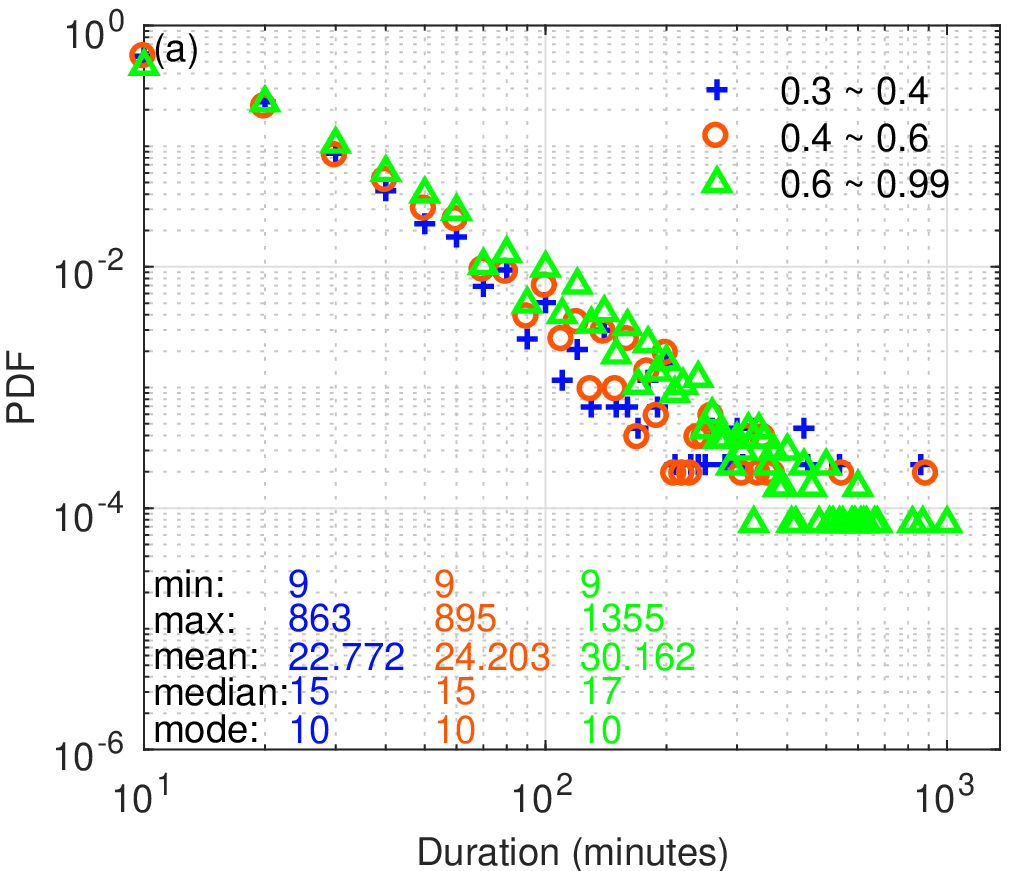}
\includegraphics[width=.48\textwidth]{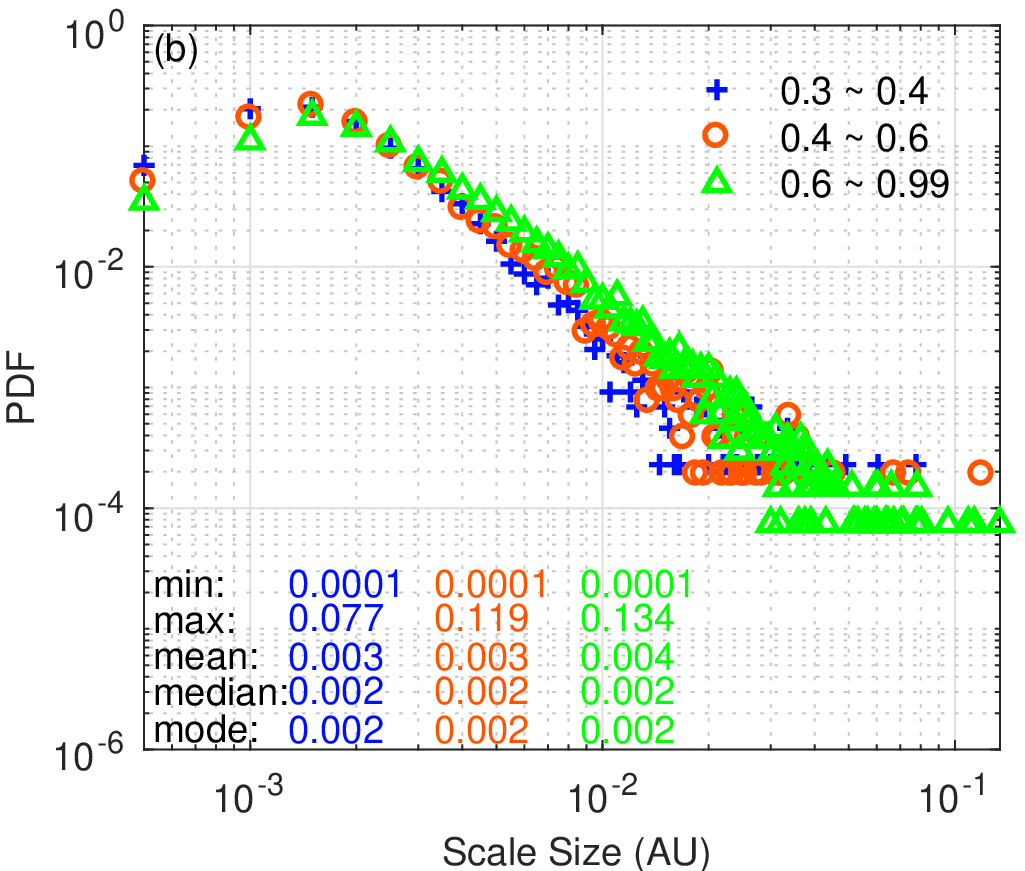}
\caption{The (a) duration and (b) scale size distributions of SFRs for
Helios 1 \& 2. The events from both spacecraft are divided into
three  groups according to the radial distances in au, as
indicated by the legend. The statistical quantities for each
distribution are also denoted in each panel.}\label{fig:sizeH}
\end{figure}
We have reported the comprehensive analysis of SFR databases
generated  for the Wind/ACE and Ulysses spacecraft missions
before. Here we have applied the automated detection algorithm to
the Helios and Voyager missions for the first time. The Helios
mission, consisting of two identical probes, took an elliptical
orbit deep into the inner heliosphere around the Sun with a
perihelion around 0.3 au. Thus it had provided the in-situ
measurements of solar wind parameters in the range of heliocentric
distances between $\sim$ 0.3 and 0.99 au over the whole mission
spanning years 1975 to 1984. We found a total number of 15,041 SFR
intervals  for Helios 1 and 7,981 for Helios 2, respectively, with
the search window size (or duration) ranging from 9 to 2255
minutes. Figure~\ref{fig:sizeH} shows the distributions in terms of the probability density functions (PDFs) of the
duration and the corresponding cross-section scale sizes of
identified SFRs, taking into account the orientation of each
identified cylindrical flux rope, relative to the spacecraft path.
They are divided into three groups according to the ranges of
radial distances in au as indicated by the legend. The maximum
duration rarely exceeds 1000 minutes, due to the relatively
frequent occurrence of data gaps. The distributions of duration
for smaller values, i.e., $<$ 100 minutes, appear to exhibit power
laws. Such trends seem to only persist for scale sizes within a
narrow range $\sim$[0.002, 0.01] au.

\begin{figure}
\centering
\includegraphics[width=.48\textwidth]{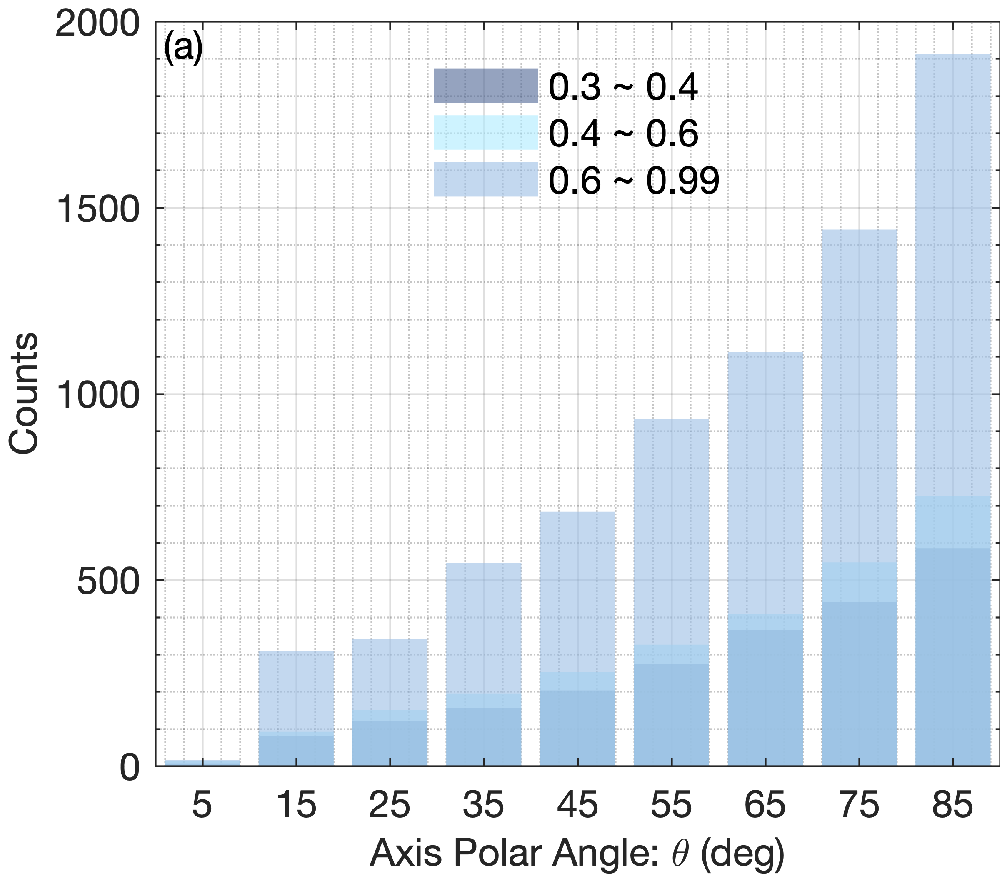}
\includegraphics[width=.48\textwidth]{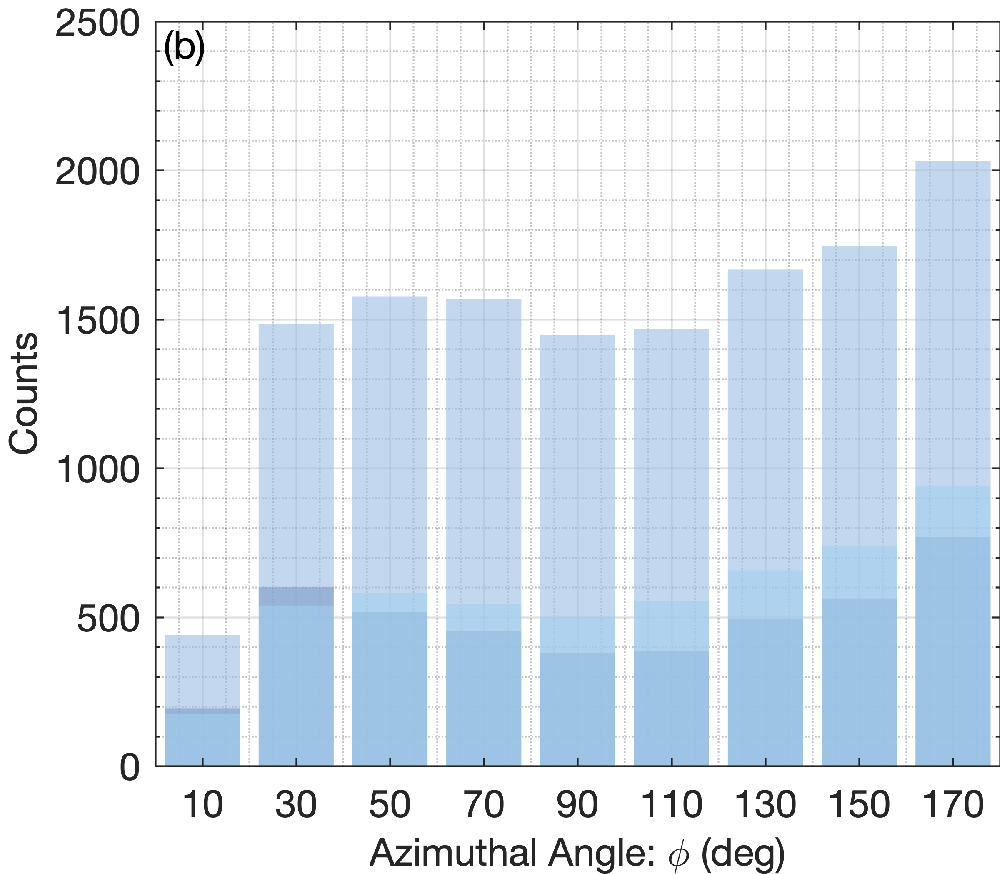}
\caption{The distributions of the directional angles of the flux
rope axis from the Helios event set: (a) the polar angle $\theta$
in degrees, and (b) the azimuthal angle $\phi$ in degrees, as
measured in the spacecraft centered $RTN$ coordinates. Different
shades represent the subgroups of events separated by the
corresponding radial distances in au as indicated by the legend in
panel (a).}\label{fig:anglesH}
\end{figure}
Figure~\ref{fig:anglesH} shows the distributions of orientation
angles of the flux rope cylindrical axis $z$ for the three groups,
separately. Figure~\ref{fig:anglesH}a shows the histograms of the
polar angle distributions overplotted for the three groups, as
indicated by the legend, corresponding to events identified at
different radial distances. They all tend to have increasing
counts toward $\theta\approx 90\deg$, i.e., increasingly more
events with smaller inclination angles with respect to the
ecliptic plane. Figure~\ref{fig:anglesH}b shows the corresponding
distributions of the azimuthal angles, $\phi$, measured with
respect to the positive $R$ axis for the $z$ axis projected onto
the $TN$ plane. The distributions are folded into the range
[$0\deg$, 180$\deg$] such that this angle measures the smaller
angle between the projected $z$ axis onto the $TN$ plane and the
positive $R$ axis. Unlike the clear tendency in the polar angle
distributions, the azimuthal angle $\phi$ has a less clear
tendency and is more broadly distributed, although it still tends
to peak near one end about 170$\deg$ and the other around
30-50$\deg$. This behavior is likely owing to the wide range of
radial distances where these event were identified. A strict
Parker spiral angle distribution would correspond to azimuthal
angles changing from near 0$\deg$ or 180$\deg$ near the Sun to
about 45$\deg$ or $135\deg$ near 1 au. Such a trend is somewhat
embedded in Figure~\ref{fig:anglesH}b.

\begin{figure}
\centering
\includegraphics[width=.48\textwidth]{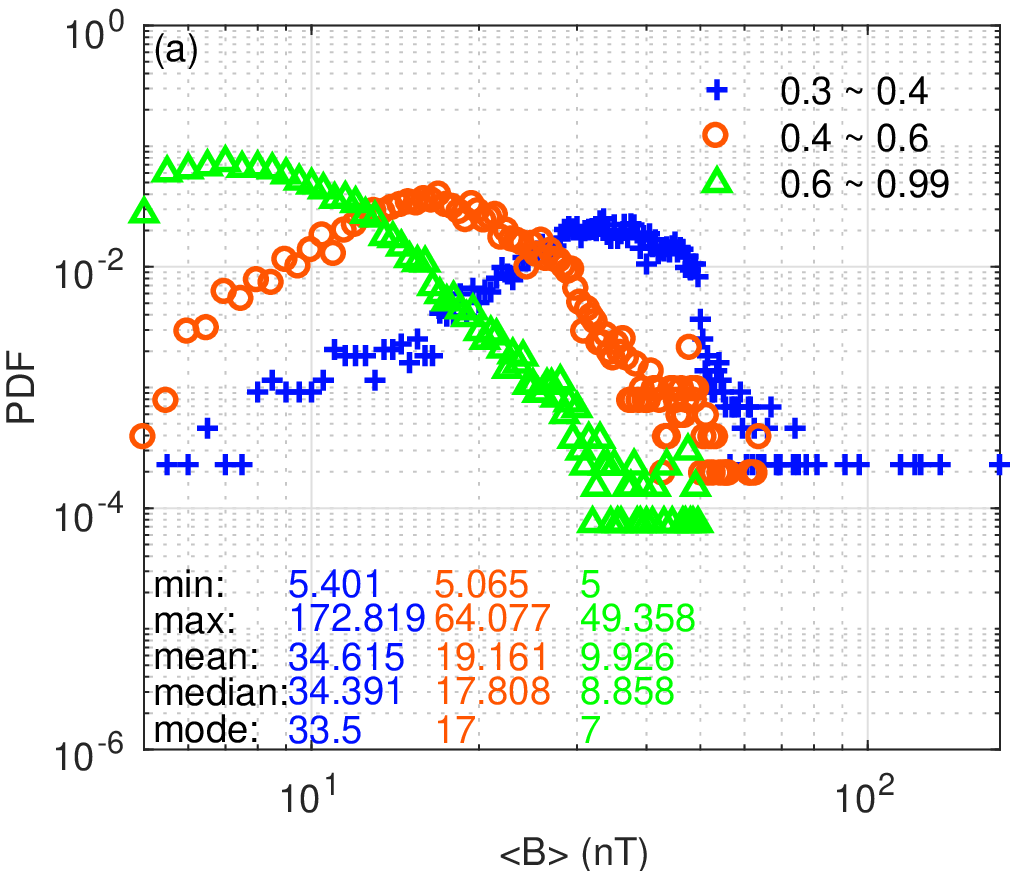}
\includegraphics[width=.48\textwidth]{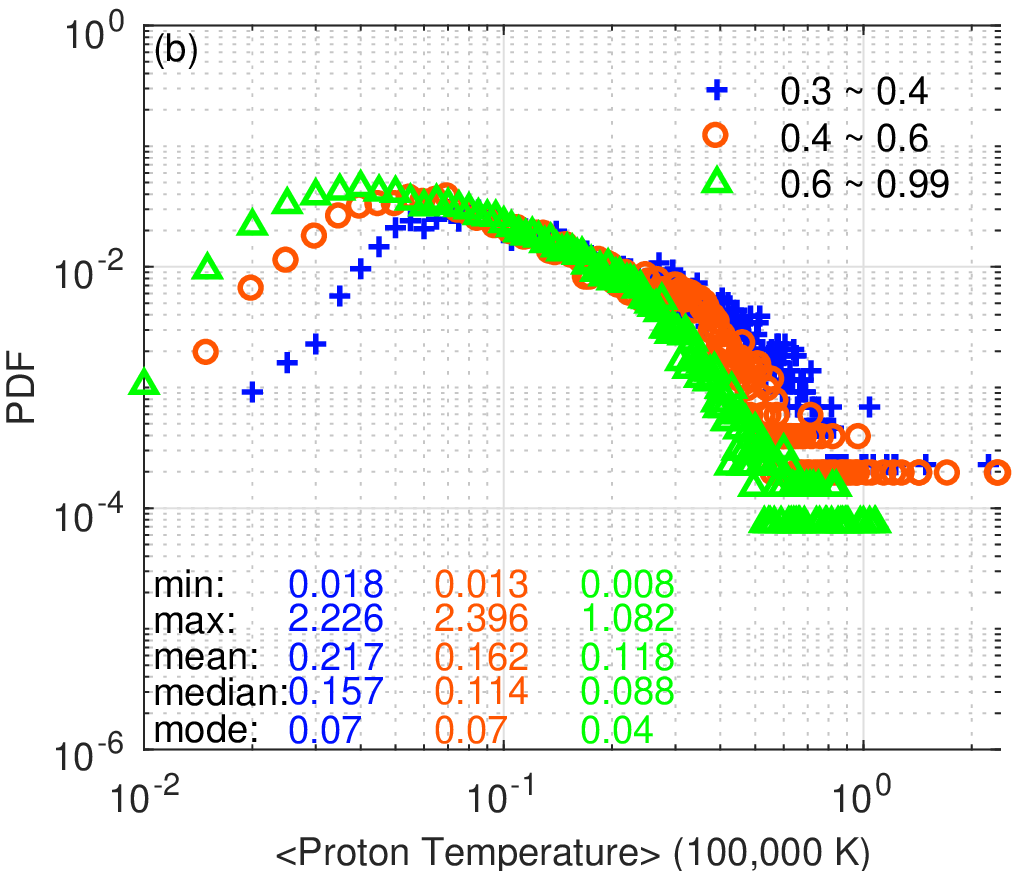}
\includegraphics[width=.48\textwidth]{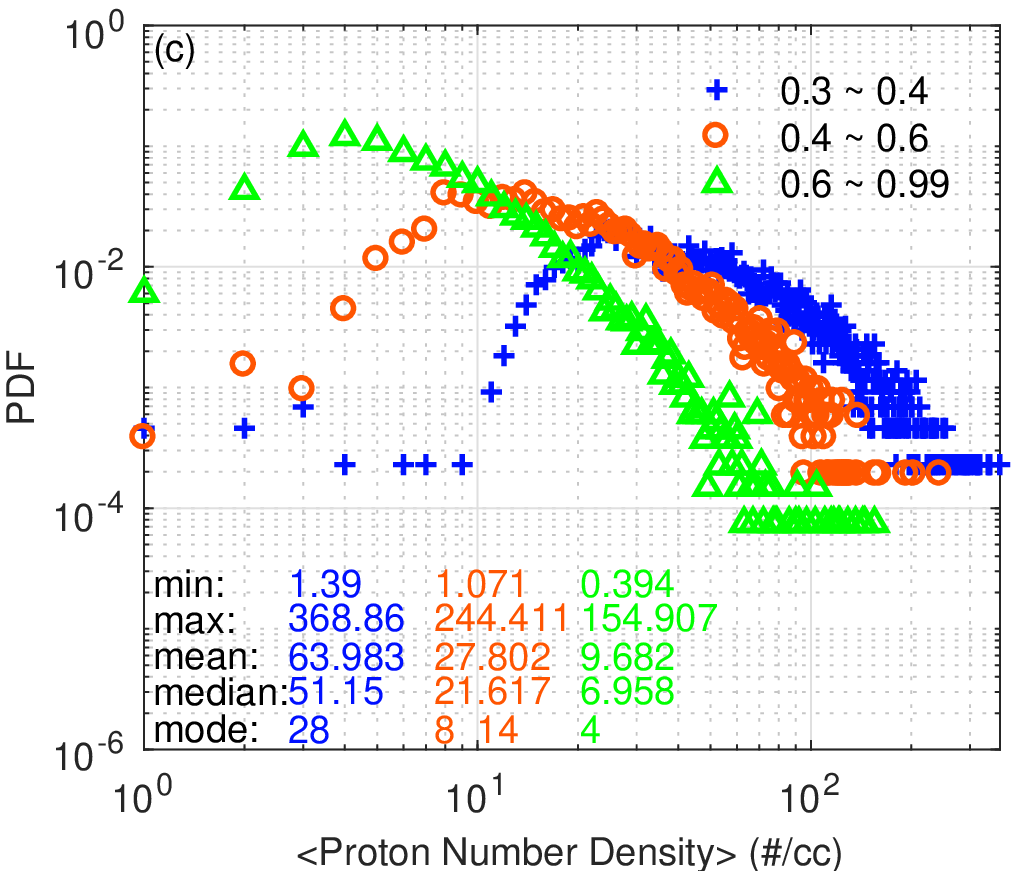}
\includegraphics[width=.48\textwidth]{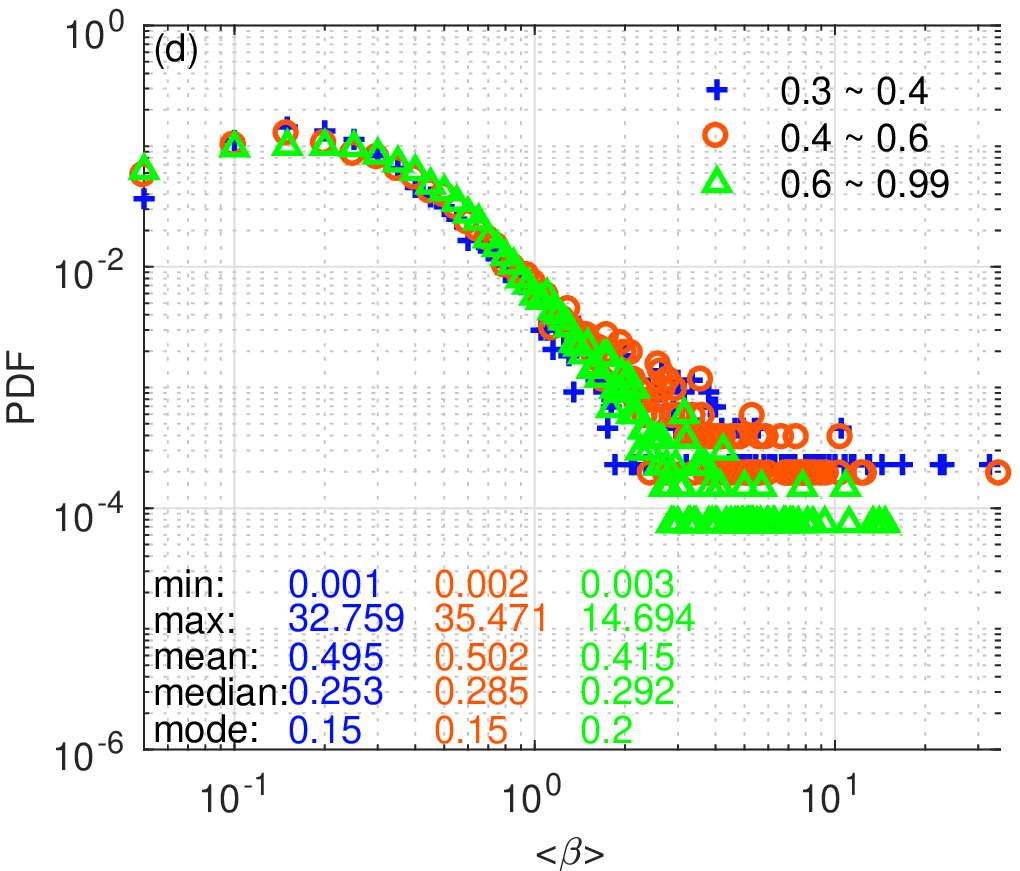}
\caption{The distributions of flux rope properties derived from
each flux rope interval identified from both Helios 1 \& 2: (a)
the average field magnitude, (b) the average proton temperature,
(c) the average proton number density, and (d) the average proton
$\beta$. Different symbols represent the corresponding subgroups
of SFR events identified with the radial distance range as
indicated by the legend. The statistical quantities for each
distribution are also denoted in each panel.}\label{fig:parH}
\end{figure}
Figure~\ref{fig:parH} shows the distributions of various flux rope
properties averaged  over each flux rope interval for all Helios
events, again separated into the three groups as before. The
radial decay in the average field magnitude $\langle B\rangle$,
and in proton number density are readily seen. On the other hand,
the changes in proton temperature and the resulting proton $\beta$
are  less pronounced. Especially for the two groups with radial
distances in the range $<0.6$ au, the distributions are nearly
identical. For all three groups, the mean proton $\beta$ values
are all around 0.5.

\begin{figure}
\centering
\includegraphics[width=.48\textwidth]{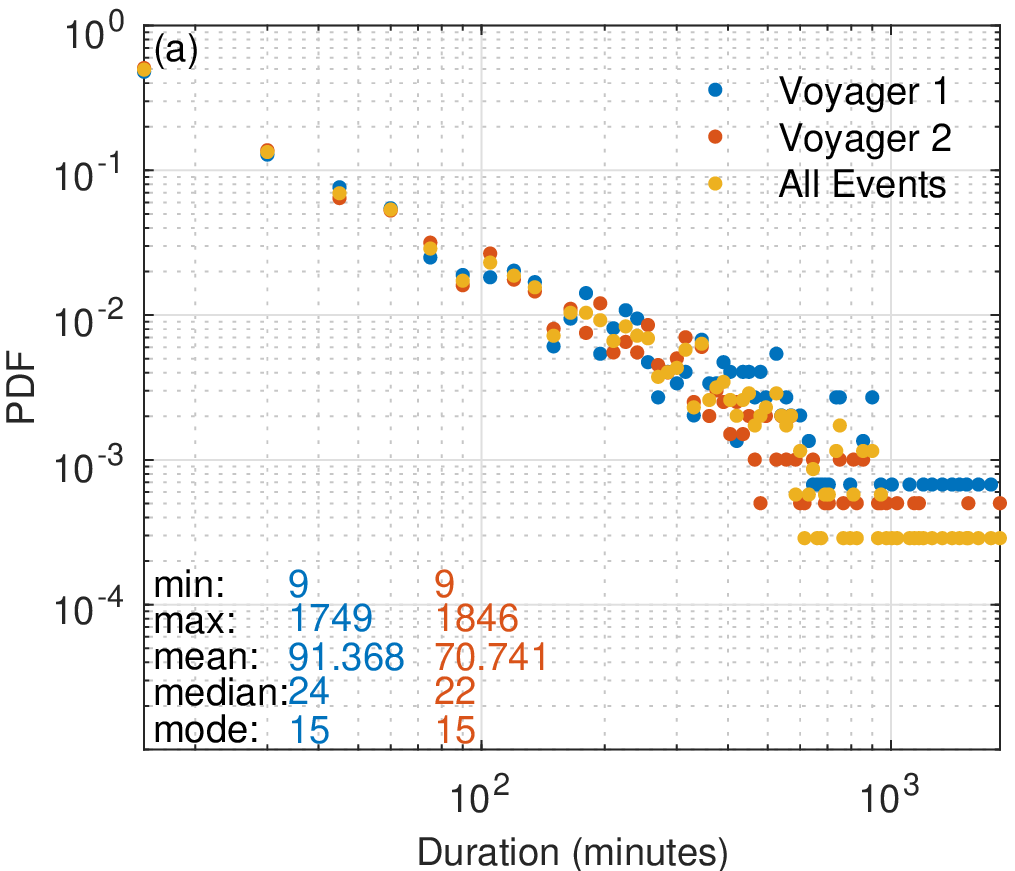}
\includegraphics[width=.48\textwidth]{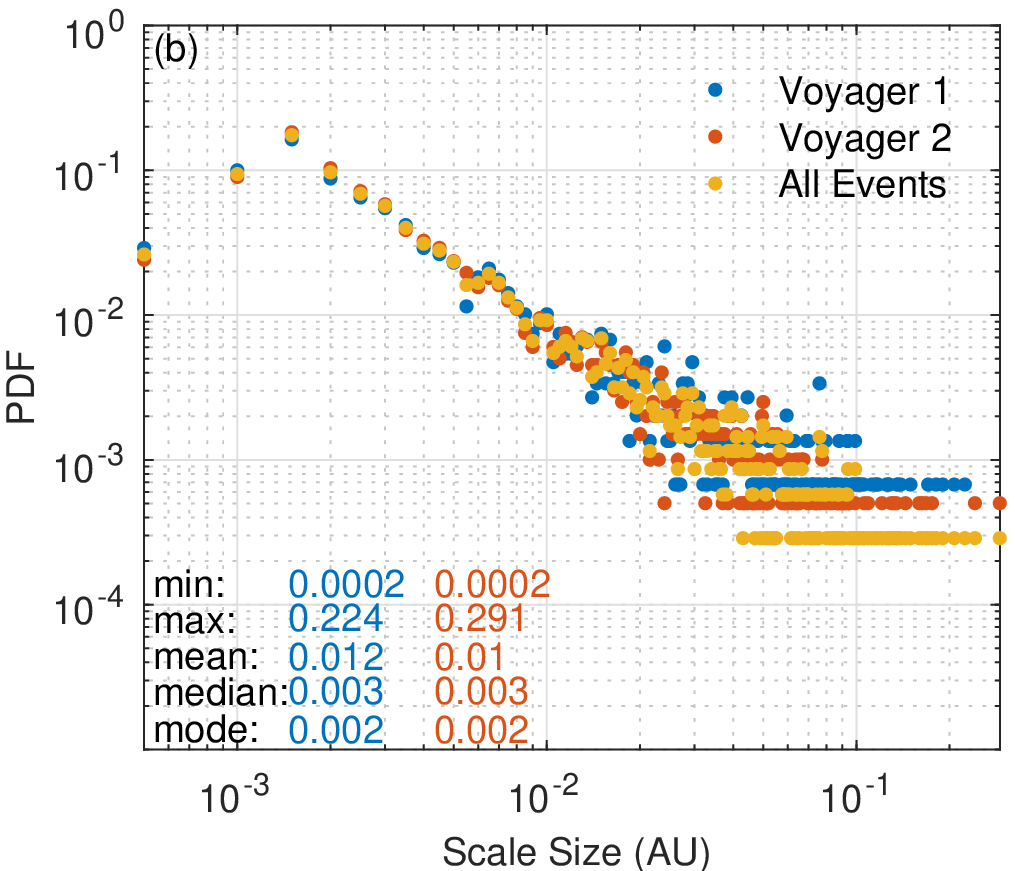}
\caption{The distributions of SFR duration and scale size for
Voyager event set. Format is the same as Figure~\ref{fig:sizeH}.
See legend for different groups.
}\label{fig:sizeV} 
\end{figure}
\begin{figure}
\centering
\includegraphics[width=.48\textwidth]{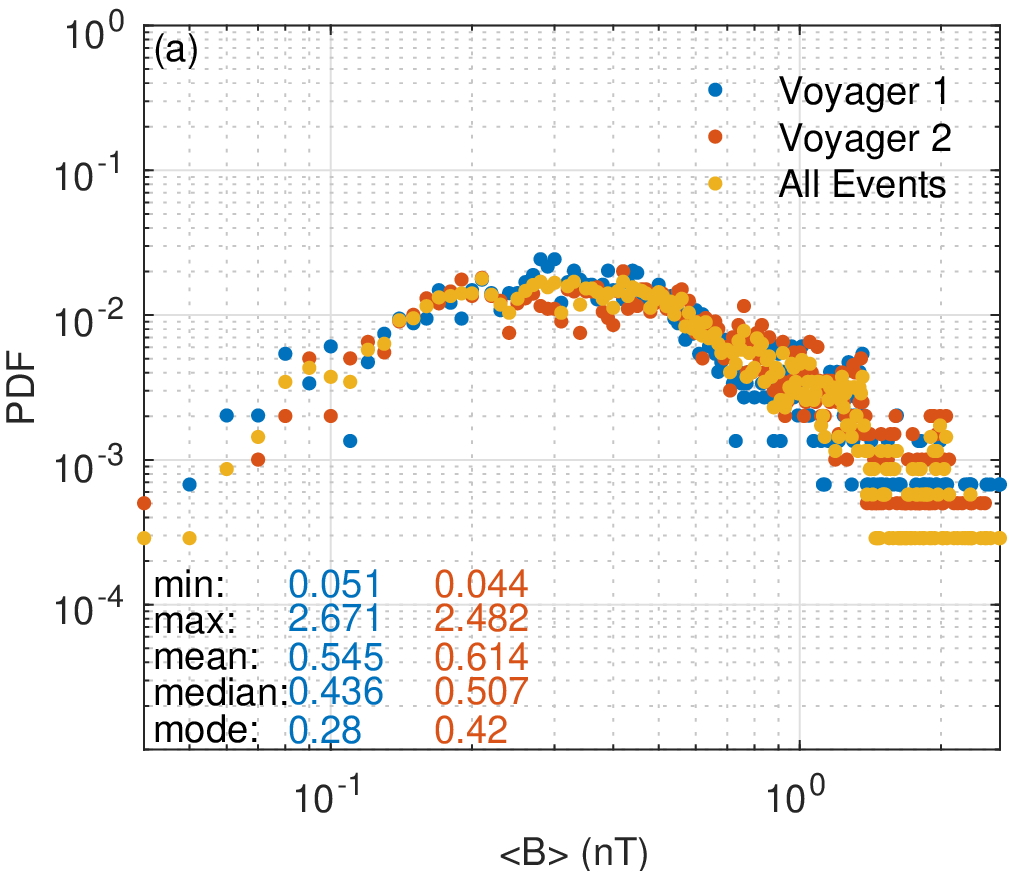}
\includegraphics[width=.48\textwidth]{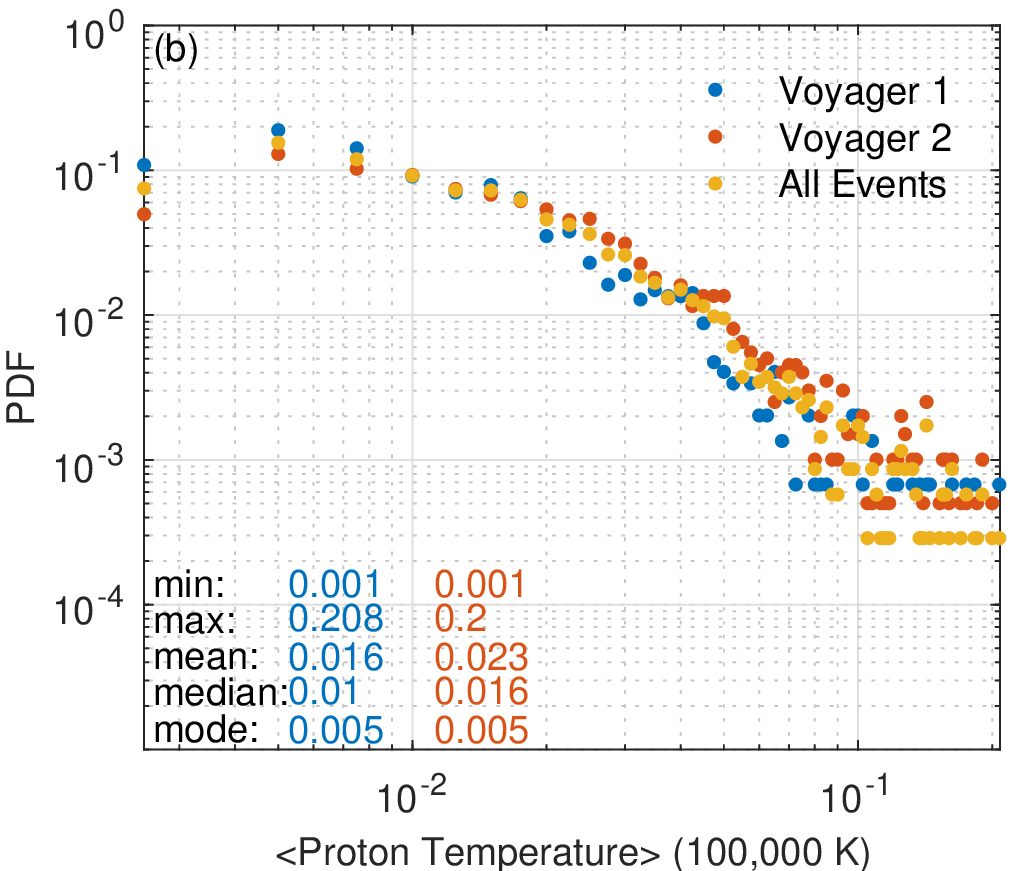}
\includegraphics[width=.48\textwidth]{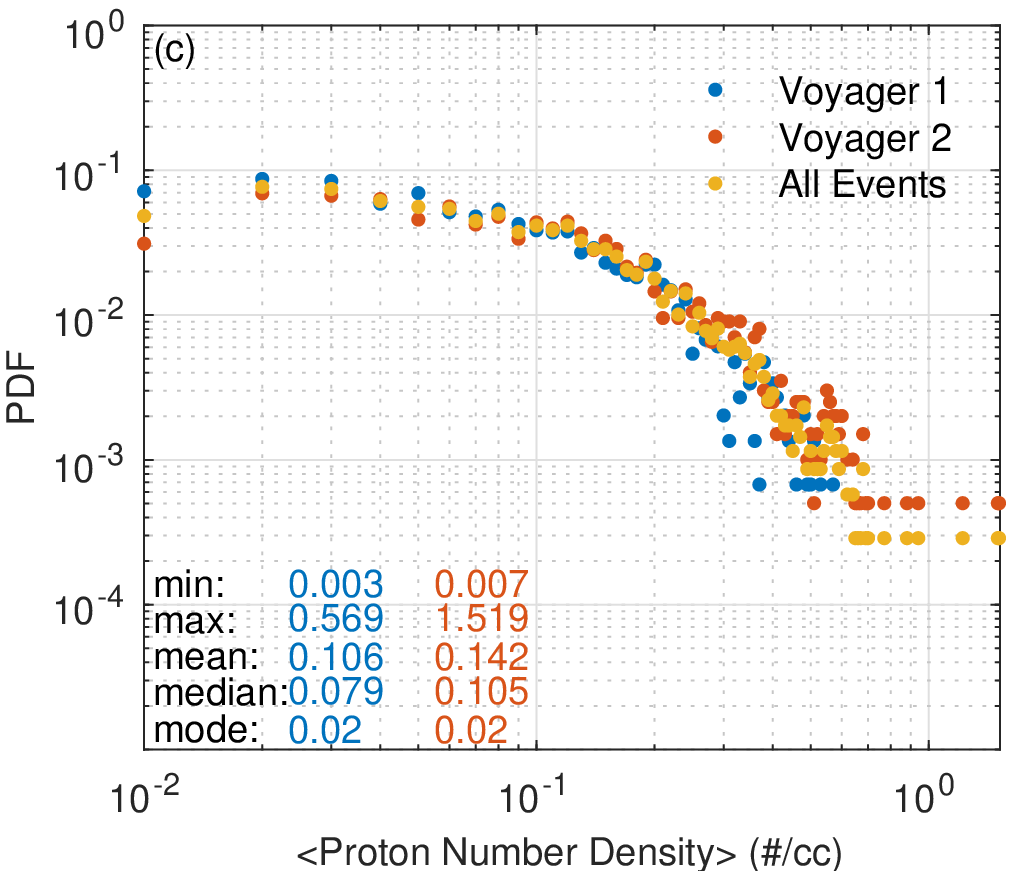}
\includegraphics[width=.48\textwidth]{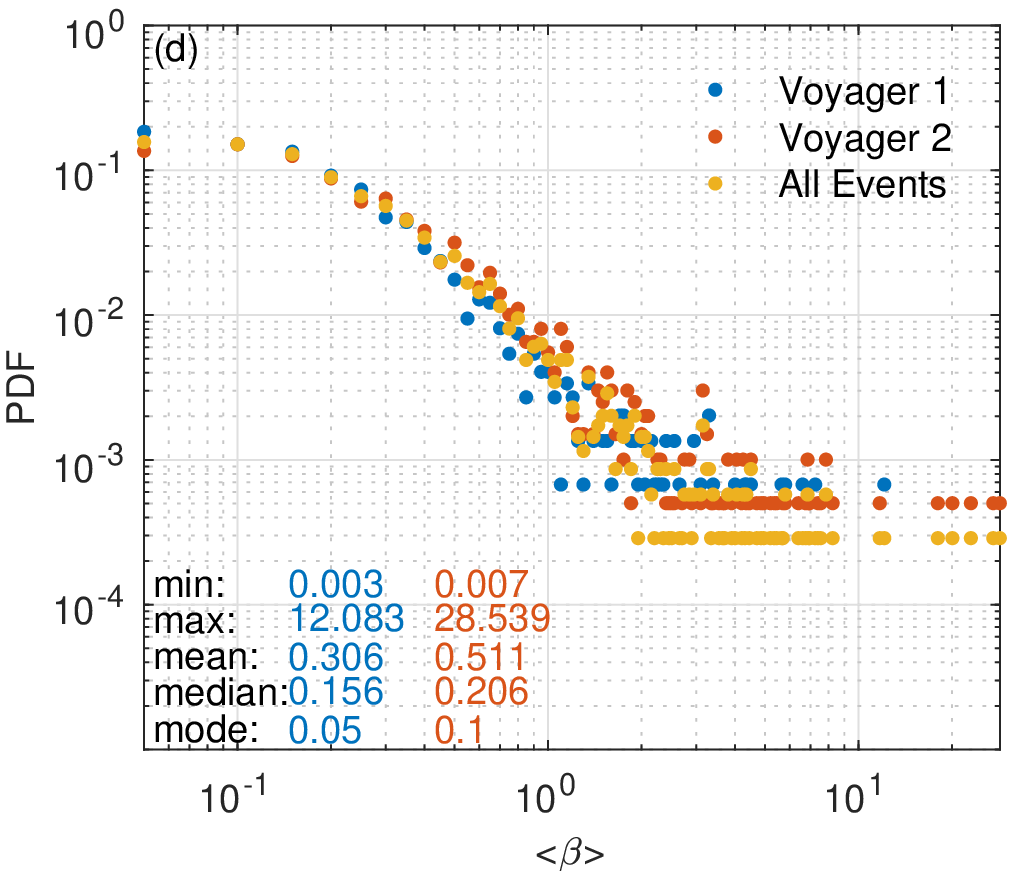}
\caption{The distributions of SFR properties for Voyagers. Format
is the same as Figure~\ref{fig:parH}. }\label{fig:parV}
\end{figure}
\begin{figure}
\centering
\includegraphics[width=.48\textwidth]{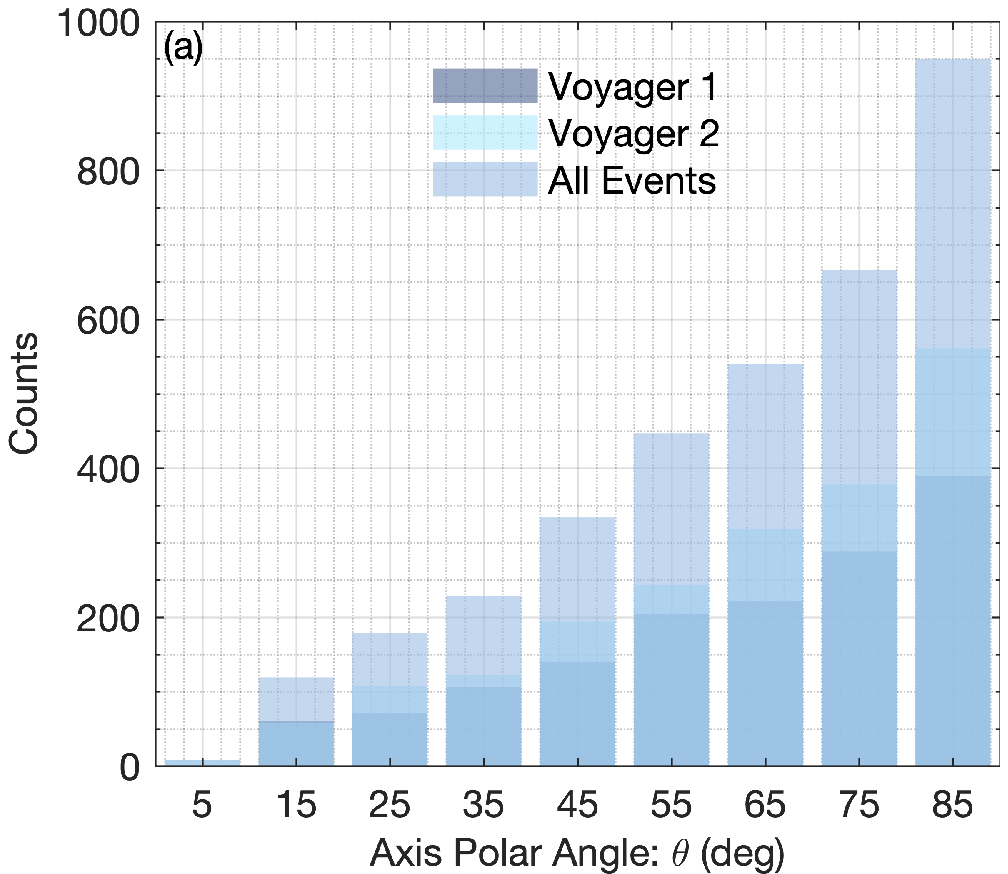}
\includegraphics[width=.48\textwidth]{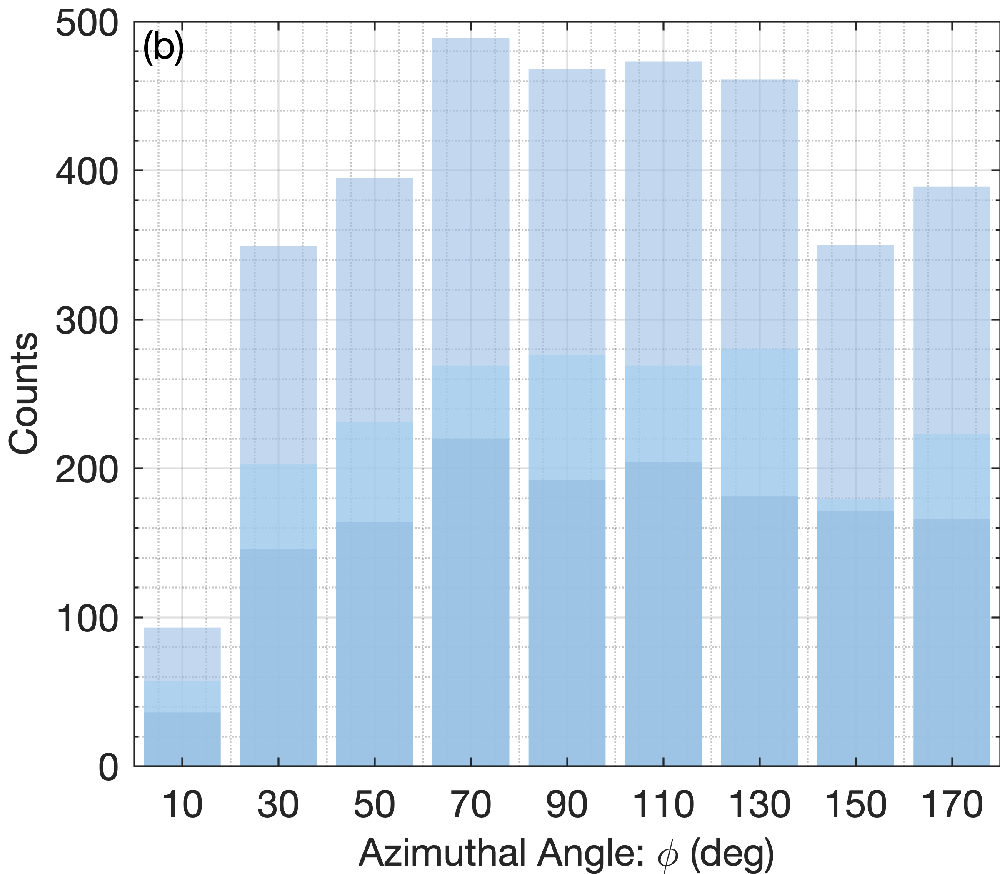}
\caption{The distributions of directional angles of flux rope axis
for Voyagers. Format is the same as Figure~\ref{fig:anglesH} (but
see the legend in a). }\label{fig:anglesV} 
\end{figure}

For Voyagers, due to the data integrity issue and the constraint
that we mainly focus on low-latitude observations in this report,
only the data in year 1980 for both Voyager 1 and 2 are analyzed.
The total number of events is very limited compared with other
missions. Therefore the statistical significance in the
distributions of relevant flux rope properties is also much
reduced. Nonetheless for completeness, we present the same set of
parameters representative of flux rope properties for the first
time at the
 heliocentric distance around and beyond 6 au near the ecliptic.
Figure~\ref{fig:sizeV} shows the distributions of duration and
scale size. The power-law trend is not clear due to the relatively
large scattering of the data points and low counts.
Figure~\ref{fig:parV} shows the corresponding distributions of
selected parameters averaged over flux rope intervals, similar to
Figure~\ref{fig:parH}. The decreases in the magnitudes of $\langle
B\rangle$, $\langle T_p\rangle$, and $\langle N_p\rangle$ are
evident at such a large radial distance away from the Sun. However
the resulting proton $\langle\beta\rangle$ remains modest, with an
order of magnitude $\sim 0.1$ on average, even at this radial
distance. The flux rope axis orientation angle distributions,
given in Figure~\ref{fig:anglesV}, indicate a trend with more
events lying on the ecliptic plane along the nominal Parker spiral
direction. The counts for the $z$ axis polar angle gradually
increase toward 90$\deg$, while the corresponding azimuthal angle
distribution has a broad peak around 90$\deg$ (the direction
perpendicular to the radial direction).

It is worth noting that we were not able to present the
statistical results of waiting time and wall-to-wall time
distributions \citep[see, e.g.,][]{zhengandhu2018} for both Helios
and Voyager missions because of wide-spread data gaps. These gaps
will interfere with the distributions of these two quantities.

\begin{figure}
\centering
\includegraphics[width=8.3cm]{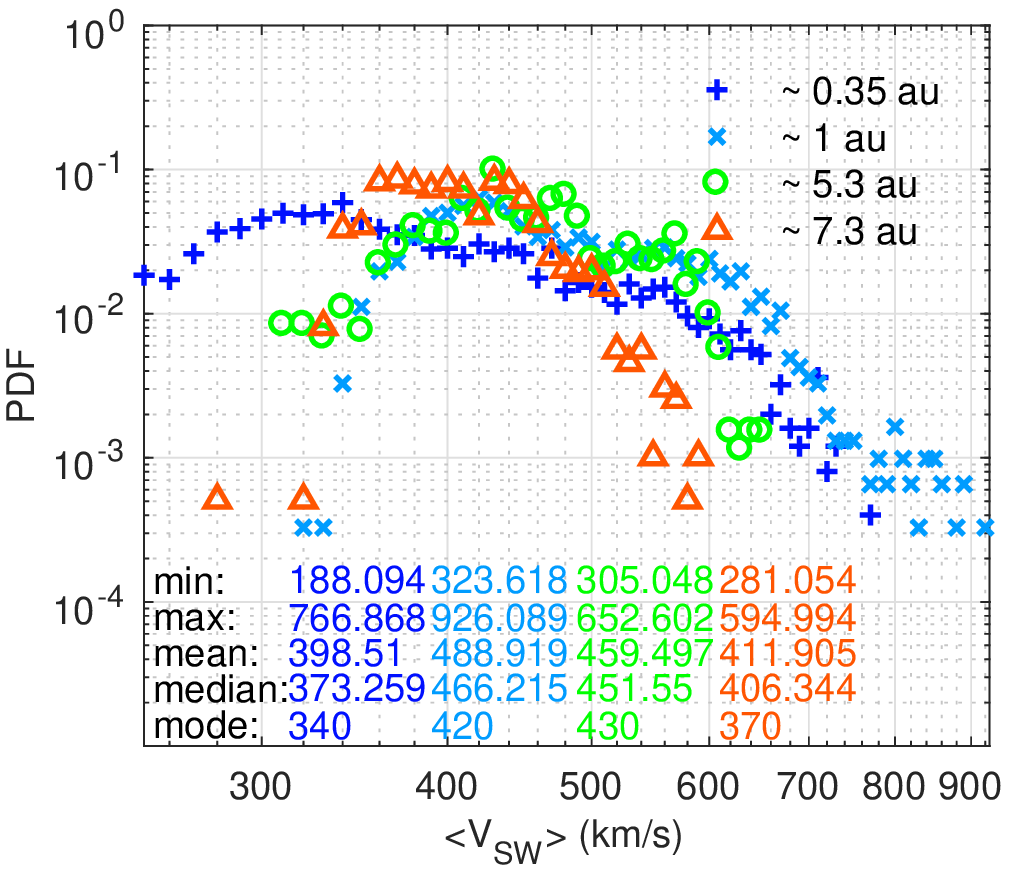}
\caption{The distributions of the average solar wind speed,
$\langle v_{sw}\rangle$, for events listed in
Table~\ref{tbl:evolution}. See legend for the correspondence of
symbols to different spacecraft missions at varying heliocentric
distances near the ecliptic. }\label{fig:vsw}
\end{figure}

\subsection{Radial Distributions of Flux Rope Properties Under Similar
Conditions}\label{sec:oneyear} To further investigate the
variation of SFR properties at different radial distances, here we
select a subset of events from each spacecraft mission. Each spans
 a time period of the same length, i.e., one year, to
facilitate intercomparison among the selected flux rope properties
under similar conditions, except for being radially distributed.
For example, Figure~\ref{fig:vsw} shows the average solar wind
velocity distributions in those flux rope intervals. The
distributions are similar if we exclude the counts in single-digit
numbers. They mostly lie in the range between 300 and 600 km/s,
with the notable exception for Helios whose counts  extend well
into the low-speed wind regime.

The selection criteria and the resulting event counts are listed
 in Table~\ref{tbl:evolution}. In a previous study by  \citet{2018ApJS..239...12H}, the possible
effect of radial distances on SFRs via ACE and Ulysses
measurements was reported. Now, with more available databases we
plan to further extend that study by selecting specific years for
four in situ datasets. Specifically, in Table \ref{tbl:evolution},
we present the basic information and special criteria applied to
these four sets of measurements: (1) For Helios, we select SFR
records detected in less than 0.4 au for both spacecraft and set a
lower limit of magnetic field magnitude  to be 25 nT to remove
small fluctuations. (2) For ACE \& Ulysses, the year 2004 is
selected since it is one of the years when Ulysses was at low
latitudes (less than 30$^{\circ}$) and at far radial distances.
(3) For Voyagers, Year 1980 is selected. Specifically, we select a
three-month period when Voyager 1 traveled from 6.9 to 7.58 au,
and a nine-month period when Voyager 2 traveled from 6.5 to 8.08
au. Then we combine these two time periods together to yield
detection result for one full year. All the detection results are
obtained by using the duration range from 9 to 2255 minutes under
the scenario with thermal pressure included in the calculation
except for Ulysses \citep{Chen2019}. Due to the 4-8 minute
resolution of plasma data and 1 min magnetic field data, we switch
off the thermal pressure for Ulysses in order to have a consistent
comparison with other missions.

\begin{table}[h]
    \caption{Selection Criteria for Event Sets from Each Spacecraft Mission over One Full Year.}\label{tbl:evolution}
    \centering
    \begin{tabular}{cccccc}
        \hline
        Spacecraft & Year & Radial Distance & Wal\'en Slope & $|\mathbf{B}|$ & Counts\\
                   &      &  (au)           &     Threshold &     (nT) & \\
\hline
        Helios 1 \& 2 & ...\tablenotemark{a} & 0.3-0.4 & 0.3 & $\ge 25$ & 2491\\
        ACE  & 2004 & $\sim$1.0 & 0.3 & $\ge 5$ & 3049\\
        Ulysses  & 2004 & 5.3-5.4 & 0.5 & $\ge 0.2$ & 2620\\
        Voyager 1 \& 2  & 1980 & 6.5-8.1 & 0.5 & - & 1967\\
    \hline
    \end{tabular}
    \tablenotetext{a}{Multiple years from 1975 to 1981.}
\end{table}

\begin{figure}
\centering
\includegraphics[width=.48\textwidth]{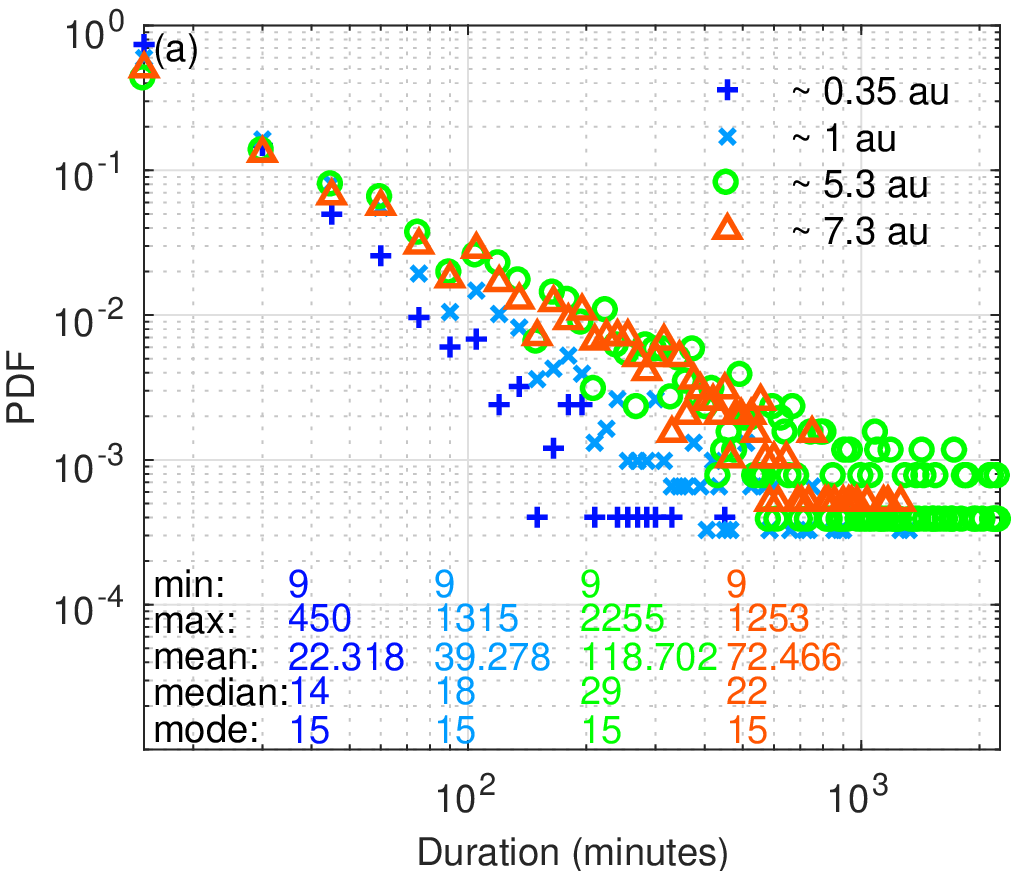}
\includegraphics[width=.48\textwidth]{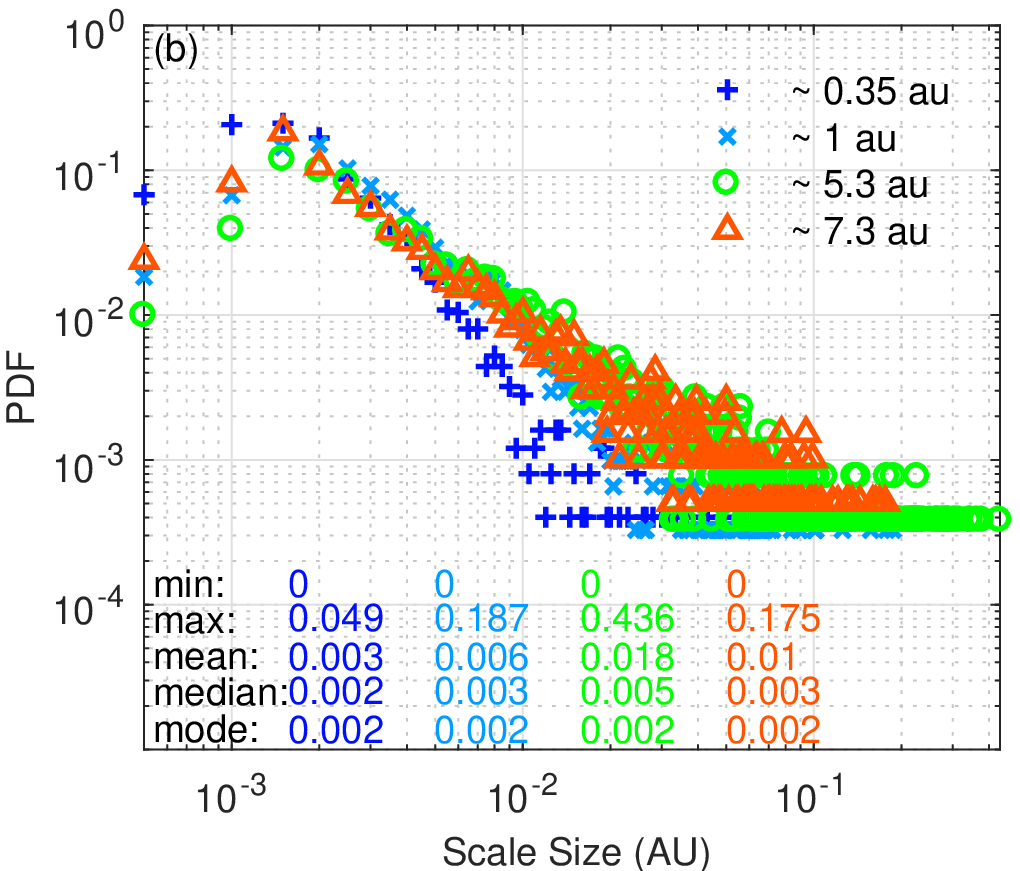}
\caption{The distributions of (a) duration and (b) scale size for
the \textbf{selected SFR} event sets listed in Table~\ref{tbl:evolution}.
Different symbols represent different missions at different radial
distances as indicated by the legend. The statistical quantities
for each distribution are denoted in each panel.}\label{fig:size}
\end{figure}

 For the Helios and Voyager
 missions, the events are selected from both probes and some from
 multiple years in order to account for the significant number of
 data gaps while maintaining a narrow range of radial distances.
 The events from ACE and Ulysses are from the same and continuous one-year period
 in 2004 when they were radially aligned near the ecliptic plane \citep{Chen2019}.
Figure~\ref{fig:size} shows the distributions of duration and
scale sizes for these selected events, respectively. Because of
reduced event counts, the power-law distribution in each event set
is not as pronounced as in each individual mission, e.g., as seen
in Figure~\ref{fig:sizeH}. They still exhibit power laws,
especially in scale size distributions shown in
Figure~\ref{fig:size}b, with different power-law slopes.  The mean
values seem to increase with radial distances, which ceases at
about 7 au at Voyagers. The disruption of such an increase at
Voyagers is likely due to the interruption in the continuous
coverage of time-series data, which prohibits the detection of
 longer/larger event intervals, although these events are much fewer.

\begin{figure}
\centering
\includegraphics[width=.48\textwidth]{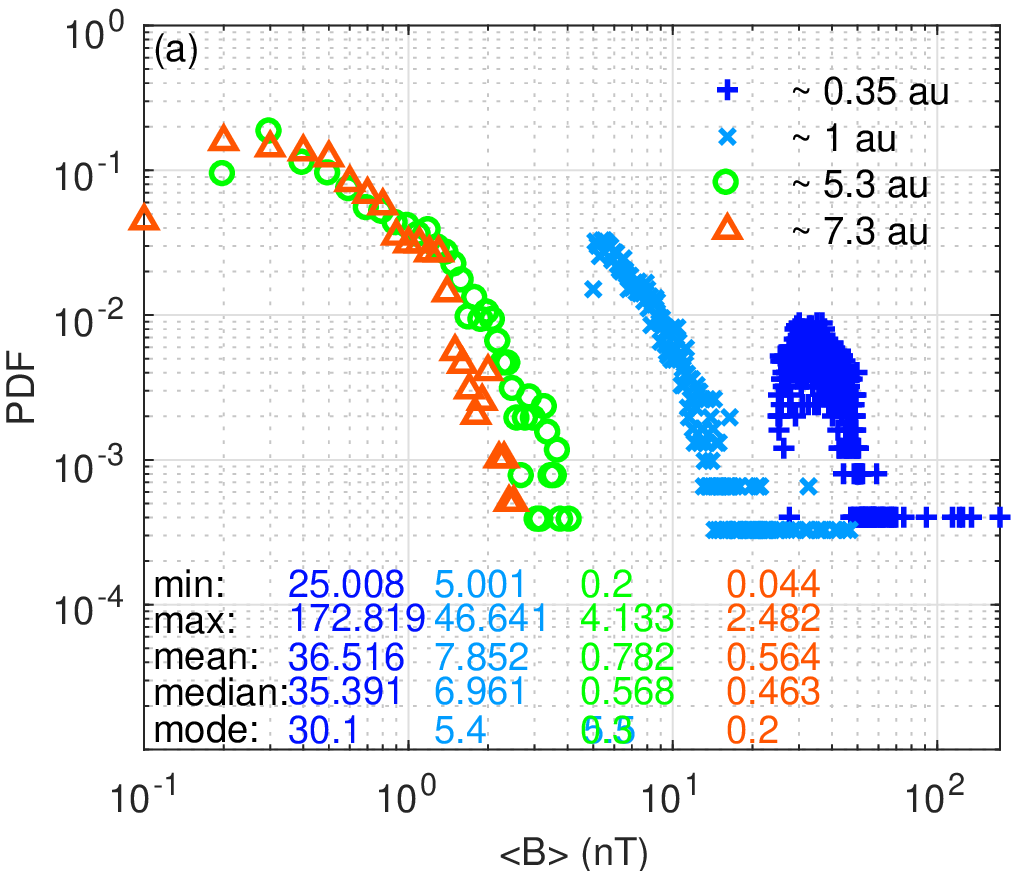}
\includegraphics[width=.48\textwidth]{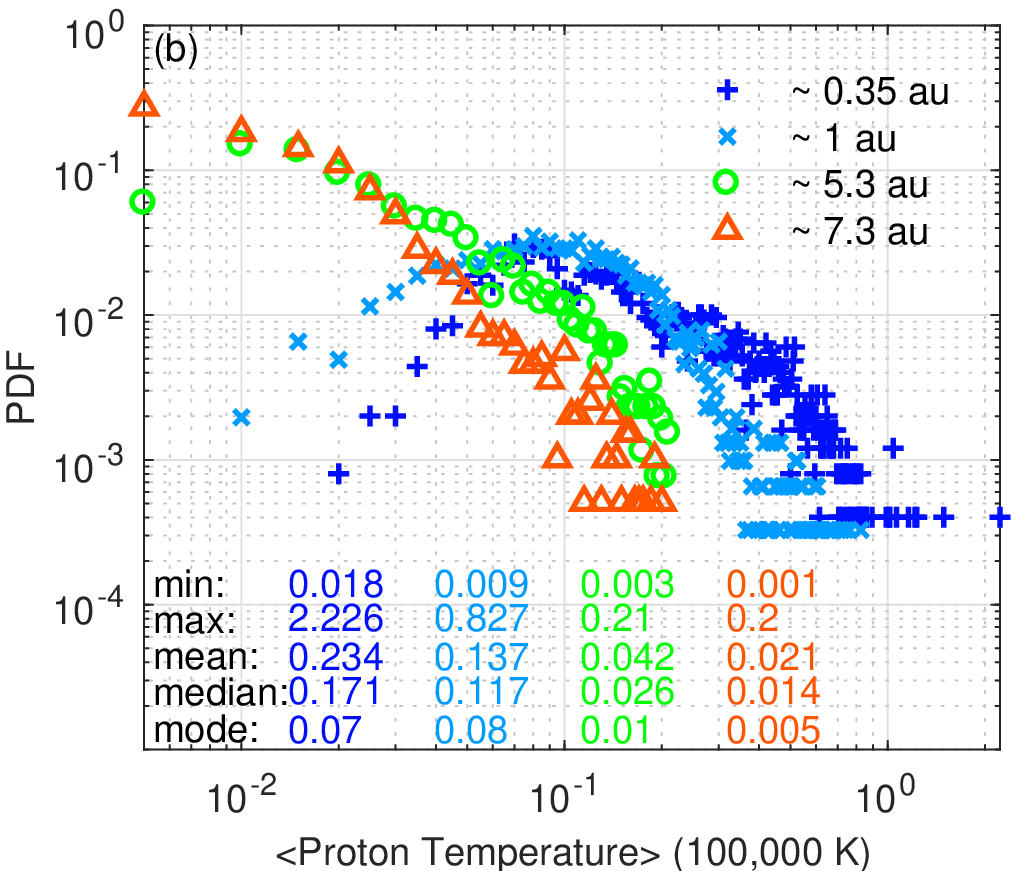}
\includegraphics[width=.48\textwidth]{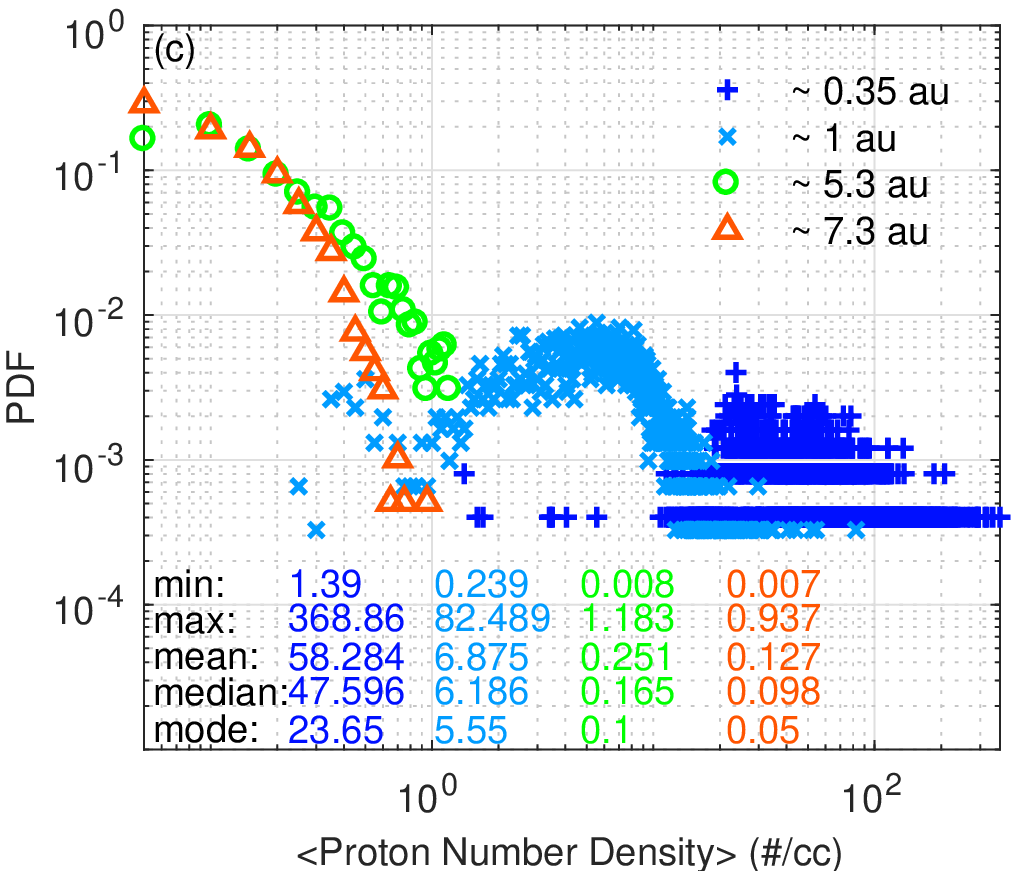}
\includegraphics[width=.48\textwidth]{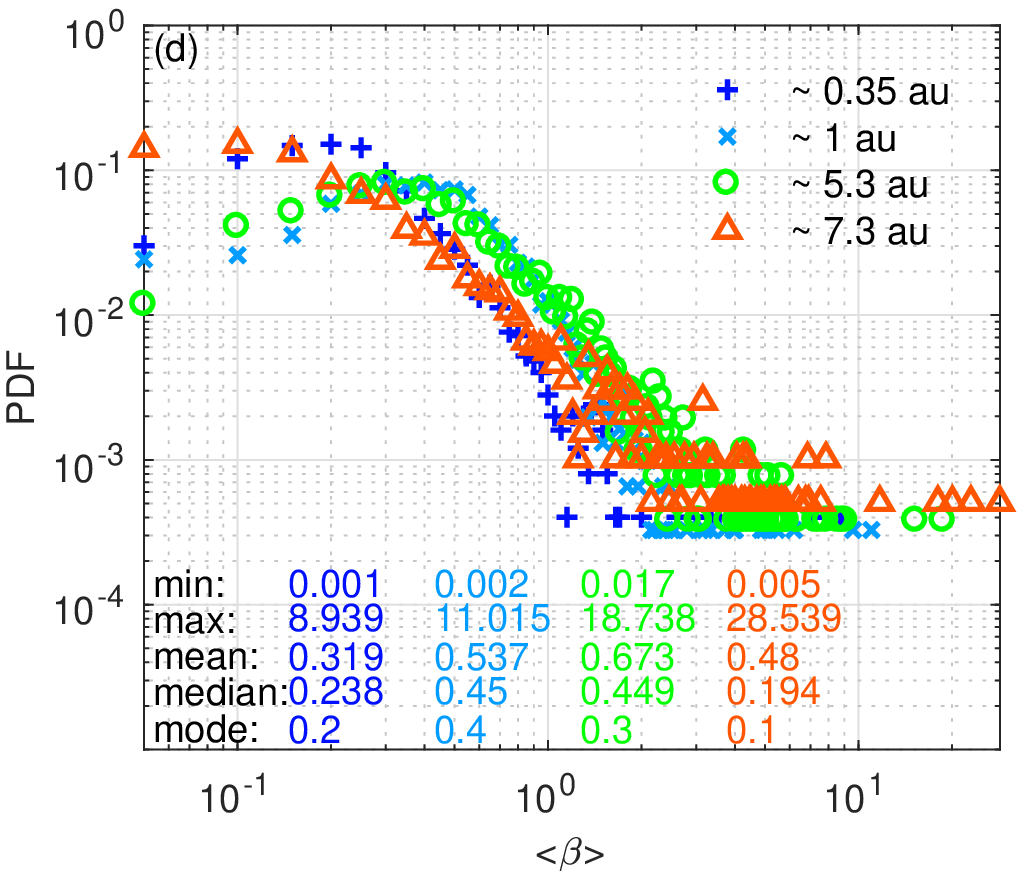}
\caption{The distributions of SFR properties for the selected
events listed in Table~\ref{tbl:evolution}. Format is the same as
Figure~\ref{fig:parH}, except that the legend indicates the
symbols for different missions at different radial
distances.}\label{fig:par}
\end{figure}

Figure~\ref{fig:par} again shows collectively the corresponding
SFR parameters. They generally exhibit variations largely owing to
the varying radial distances. Their values span wide ranges, and
are well separated for different radial distances. One exception
is the proton $\langle\beta\rangle$, which exhibits much similar
distributions among all the event sets, with the mean values of
the same order of magnitude, $\gtrsim 0.1$, despite the wide
separation in radial distances.

\begin{figure}
\centering
\includegraphics[width=.48\textwidth]{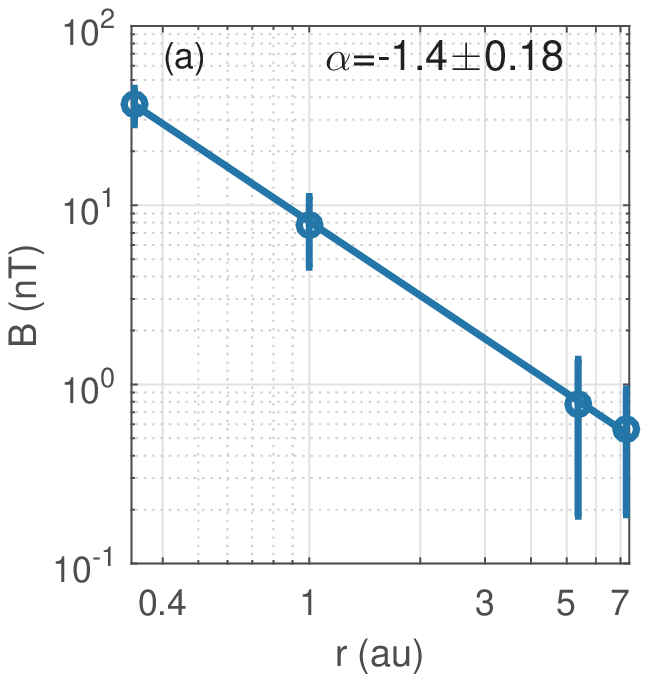}
\includegraphics[width=.48\textwidth]{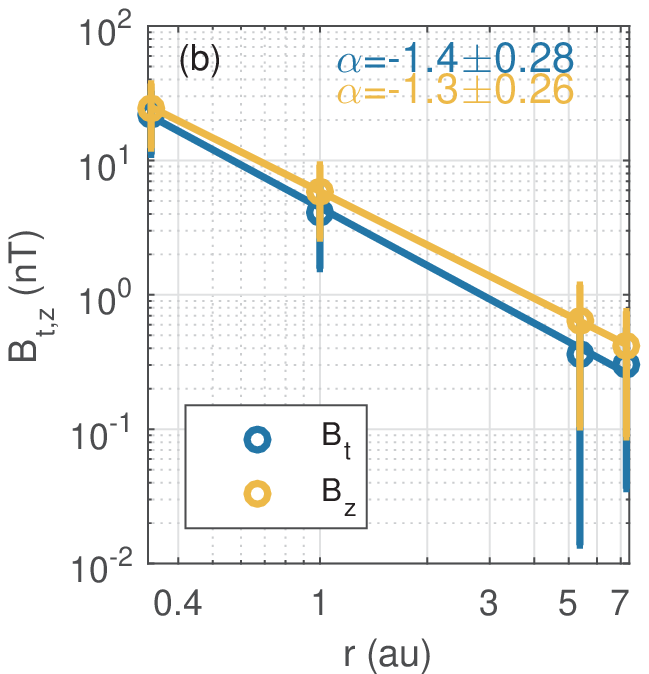}
\caption{The radial dependence with the heliocentric distance $r$
for (a) the average magnetic field magnitude, and (b) the
corresponding transverse and axial component, respectively. The
symbols with errorbars represent the averages over all selected
flux rope intervals and associated standard deviations, while the
solid lines represent the linear fits, $\propto r^\alpha$. The
corresponding slopes ($\alpha$) on the log-log scales are given.
}\label{fig:rB4}
\end{figure}

Finally, to examine and quantify the trend in radial variation for
a few selected flux rope parameters, we show the radial variations
of the average field magnitude $\langle B\rangle$, the average
transverse field $\langle B_t\rangle$, and the axial field
$\langle B_z\rangle$ over all flux rope intervals with associated
uncertainties (standard deviations) for each event set at the
corresponding radial distance $r$. A power-law fit in the form
$\propto r^\alpha$ is also obtained for each quantity and the
corresponding power-law indices are denoted, as given in
Figure~\ref{fig:rB4}. They show a consistent trend of decaying
with increasing radial distances, with the power-law indices
$\alpha\approx$ -1.4. This value falls between -1 and -2. The
former number corresponds to the radially decaying power-law index
for the azimuthal component of the nominal Parker spiral field,
while the latter corresponds to that for the radial field.

\begin{figure}
\centering
\includegraphics[width=8.3cm]{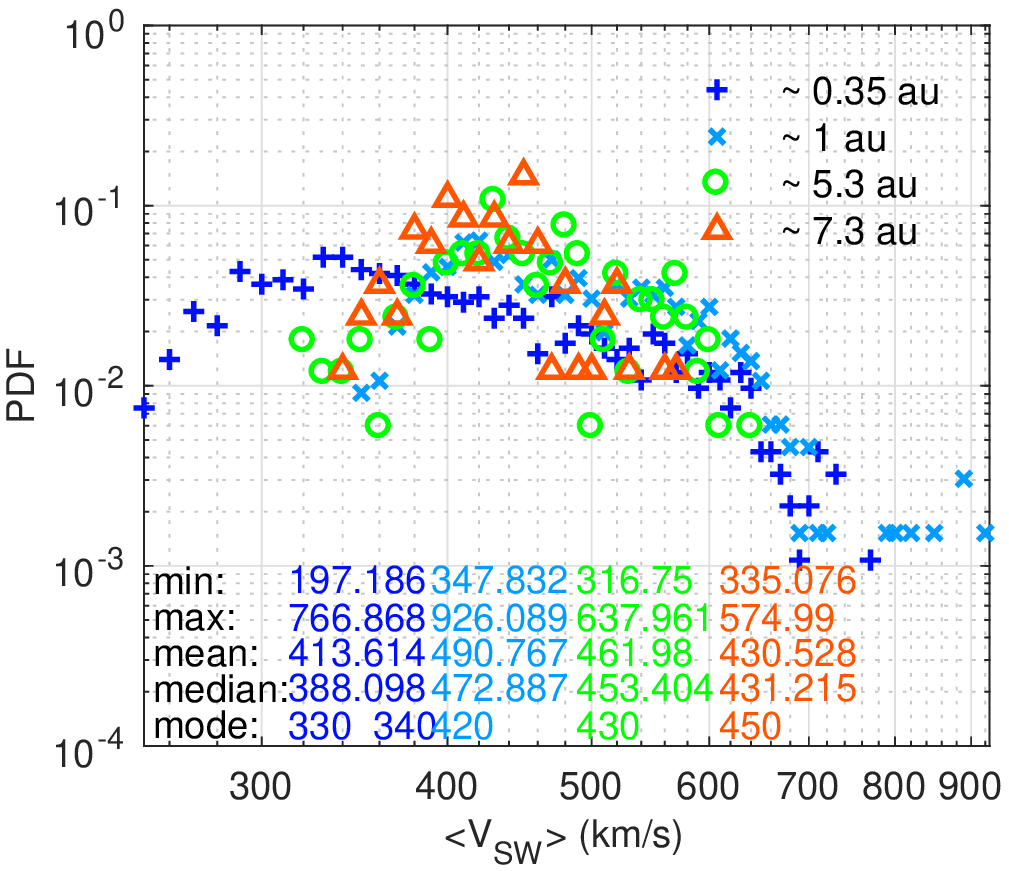}
\caption{The distribution of the average solar wind speed,
$v_{sw}$, for events listed in Table~\ref{tbl:r}. Format is the
same as Figure~\ref{fig:vsw}. }\label{fig:vsw3}
\end{figure}

\section{Radial Evolution of Magnetic Flux Ropes} \label{sec:evolution}
One scenario for the radial evolution of SFRs is to consider flux
rope merging associated with inverse magnetic energy transfer
under the assumption that the poloidal magnetic flux $A_m$ remains
conserved. We choose a typical range of poloidal magnetic flux
$A_m\in$  [0.5, 1.5] T$\cdot$ m and examine the corresponding
radial evolution of selected properties. A preliminary analysis of
the radial evolution of selected SFR properties \citep{Chen_2018}
hinted at power-law decaying relations with respect to radial
distance with power-law indices ranging from -1.5 to -0.5 for a
wide range of $A_m$ values. This quantity, $A_m$, represents the
amount of poloidal magnetic flux per unit length (per meter) for a
cylindrical flux rope, which is an output from our GS based search
algorithm and is directly derived from in-situ spacecraft
measurements. It is simply the difference between two flux
function values, one at the center and the other at the boundary
of a flux rope.

In order to compare with the relevant theoretical work
\citep[e.g.,][]{Zhou2019}, we consider a scenario of consecutive
flux rope merging, leading to an increase in scale size but a
decrease in magnetic field magnitude, while maintaining the
poloidal magnetic flux. We further scrutinize our event sets by
imposing the criterion for the unit poloidal magnetic flux, $A_m$,
to be within [0.5, 1.5] T$\cdot$m, i.e., approximately a constant
$\sim 1$ T$\cdot$m with uncertainty. Such a value around 1
T$\cdot$m is typical for an SFR in the solar wind. A quantitative
case study of flux rope merging in interplanetary space was
reported in \citet{Hu2019iop}. It was found that given a dynamic
evolution time  on the order of $10^4$ seconds for the two
adjacent flux ropes to fully merge into one via magnetic
reconnection, the reconnected magnetic flux during the process
would be equal to the amount of the poloidal flux of each flux
rope. If we take the typical value 1 T$\cdot$m, then the
reconnection rate can be estimated to be approximately 1/$10^4$
V/m or 0.1 mV/m, which is not unreasonable for the solar wind,
given that a value on the order of $\lesssim$1 mV/m was typically
found for the reconnection events at the Earth's magnetopause
\citep[e.g.,][]{2010hase}. Such a range also ensures a more certain  flux rope configuration  for the corresponding events by excluding the ones with small values of $A_m$ (i.e., $<0.5$ T$\cdot$m), which are sometimes caused by a small rotating field component ($B_y$), indicating a less certain flux rope configuration. Additionally, for the large-scale
counterparts of SFRs, i.e., the MCs, the poloidal flux can amount
to $\sim$100 T$\cdot$m, corresponding to total poloidal magnetic
flux of the order of magnitude $\sim10^{21-22}$ Mx that can be
well related to the flux contents in solar source regions
\citep{Qiu2007,Hu2014}.

\begin{figure}
\centering
\includegraphics[width=8.3cm]{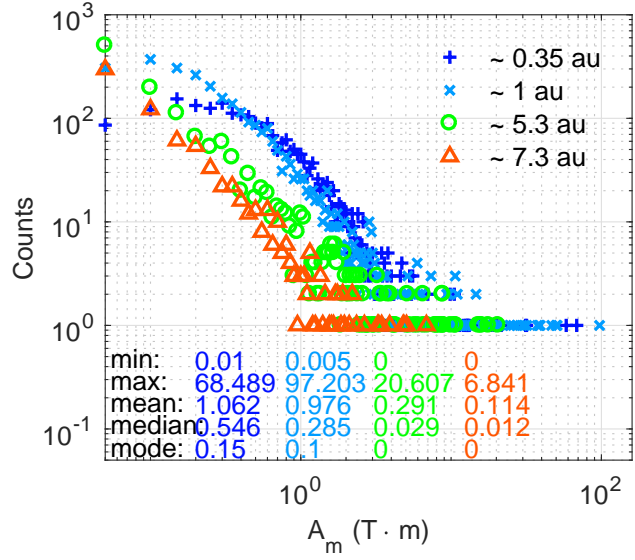}
\caption{The distributions of the poloidal magnetic flux per unit
length, $A_m$, for all the event sets listed in
Table~\ref{tbl:evolution}.}\label{fig:Am}
\end{figure}

\begin{figure}  
\centering
\includegraphics[width=.32\textwidth]{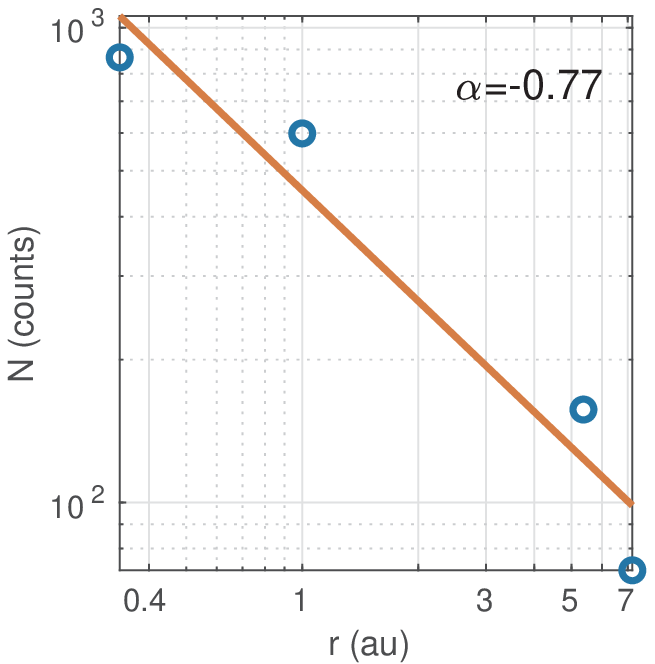}
\includegraphics[width=.32\textwidth]{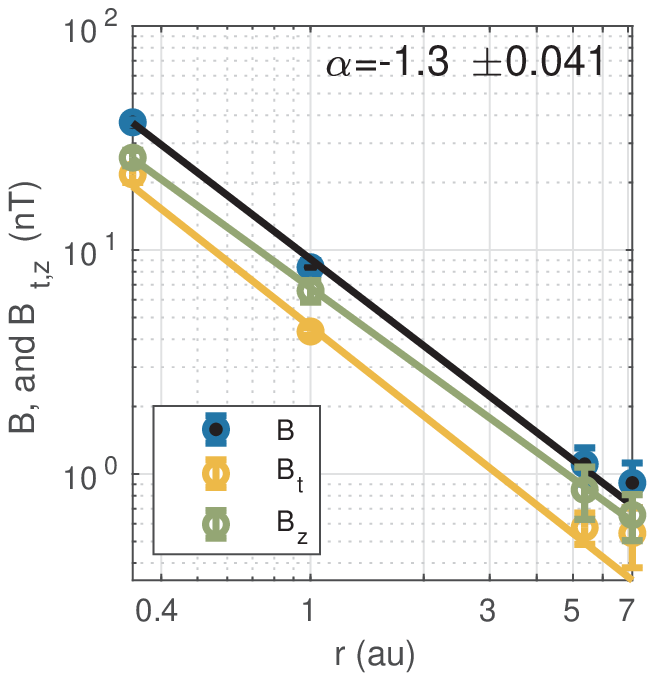}
\includegraphics[width=.32\textwidth]{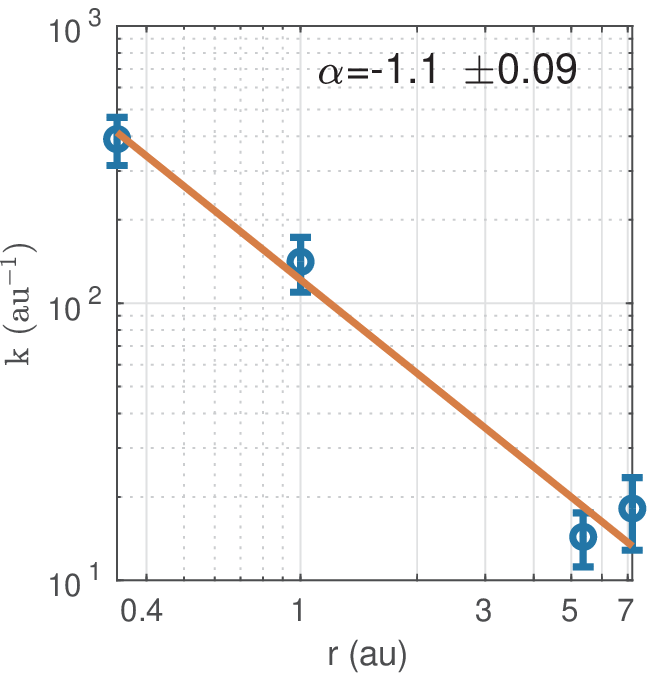}
\caption{The radial dependence of the parameters listed in
Table~\ref{tbl:r} (from the left to the right panels): the SFR
event counts, the field magnitude, and the transverse and axial
field components, respectively, and the inverse scale size. Format
is the same as Figure~\ref{fig:rB4}.  }\label{fig:r}
\end{figure}

\begin{table}[h]
    \caption{Radial distribution of the parameters with uncertainties for flux ropes with unit poloidal magnetic flux $A_m\in[0.5,1.5]$ T$\cdot$m.}\label{tbl:r}
    \centering
    \begin{tabular}{cccccc}
        \hline
        $r$ & $N$ & $\langle B\rangle\pm\Delta B$ & $k\pm\Delta k$ & $\langle B_t\rangle\pm\Delta B_t$ & $\langle B_z\rangle\pm\Delta B_z$\\
         (au)      &     &   (nT) &  (au$^{-1}$) &  (nT) & (nT) \\
\hline
0.34& 866 &37$\pm 0.91$& 391$\pm 77$& 22$\pm 1.7$ & 26$\pm 2.0$\\
    1.0    &   599& 8.4$\pm 0.59$ & 141$\pm  32$ & 4.3$\pm 0.17$ & 6.6$\pm
    0.72$\\
5.4 & 157 & 1.1$\pm 0.20$ & 14$\pm 3.2$ & 0.58$\pm
0.092$ & 0.85$\pm 0.22$ \\
7.2 & 72 & 0.91$\pm0.21$ & 18$\pm5.3$ & 0.54$\pm0.16$ & 0.66$\pm
0.15$\\
    \hline
    \end{tabular}
\end{table}
Figure~\ref{fig:vsw3} shows the corresponding solar wind speed
distributions for these subsets of events. Again there are a
significant number of events from Helios with relatively low speed
around 300 km/s. Figure~\ref{fig:Am} shows the distributions of
the poloidal magnetic flux $A_m$ for the four spacecraft missions
at different radial distances. They generally span a similar range
with the maximum values in the order of tens of T$\cdot$m, except
for the Voyagers. The event counts generally decrease with
increasing radial distances, especially for $A_m\gtrsim$ 0.5
T$\cdot$m. In \citet{Zhou2019}, it was predicted that the time
evolution of consecutive flux rope merging yields the following
power laws in time, $t$,
\begin{equation} \bar{N}\sim t^{-1.0}, \quad  \langle B\rangle \sim
t^{-0.5}, \quad  k\sim t^{-0.5}.\label{eq:t}
\end{equation}
These quantities are the flux rope count $\bar{N}$ over a 2D
domain, the average magnetic field magnitude $\langle B\rangle$,
and the inverse scale size $k$, under the condition that poloidal
magnetic flux is conserved after merging. In other words, the
amount of poloidal magnetic flux contained in each individual flux
rope is not varying in time, although two neighboring flux ropes
will merge into one consecutively. We intend to carry out an
analysis from observations and the GS-based search result,
following these conditions, with the additional assumption that
the evolution in time as envisaged by \citet{Zhou2019} can be
translated into the variation in radial distance, $r$, as we
demonstrate below, by considering a constant radial solar wind
flow. For instance, in Figure~\ref{fig:vsw3}, the mean values
among the average solar wind speed distributions for these events
are similar. One obvious caveat that is also inevitable in this
type of analysis of spacecraft data is the intrinsic radial change
of the background field, i.e., the Parker interplanetary field in
the solar wind, which was not incorporated in the theory of
\citet{Zhou2019}.

Table~\ref{tbl:r} summarizes the findings from the analysis of our
unique event sets, with values of $A_m$ strictly confined within
the range $[0.5,1.5]$T$\cdot$m. The list of parameters includes
the event counts $N$, $\langle B\rangle$, $k$, $\langle
B_t\rangle$, and $\langle B_z\rangle$ with associated
uncertainties at the corresponding radial distances, $r$. These
results are further illustrated in Figure~\ref{fig:r} with the
corresponding power-law fittings, $\propto r^\alpha$. The count
$N$ is obtained only along the radial dimension. In order to
compare with the count predicted by \citet{Zhou2019} from their 2D
simulations over a square area, we may compute $N^2$ to
approximate such a count in 2D, i.e., $N^2\approx\bar{N}$. This
yields a power-law index $\alpha\approx -1.5$. Therefore, similar
to what \citet{Zhou2019} obtained as given in
equation~(\ref{eq:t}), the radial evolutions of these quantities
seem to obey the following power laws from our data analysis,
\begin{equation} N^2\sim r^{-1.5}, \quad \langle B\rangle \sim
r^{-1.3}, \quad  k\sim r^{-1.1}.\label{eq:r}
\end{equation}
This set of power laws still differs significantly from those
presented by \citet{Zhou2019}, when assuming the equivalence
between the radial distance $r$ and time $t$.

\section{Summary and Discussion} \label{sec:discussion}
In summary, we have carried out quantitative analysis of the
interplanetary spacecraft mission data, in addition to the
ACE/Wind and Ulysses missions, following the automated detection
approach for the SFRs based on the GS reconstruction technique.
The new results reported here yield the following total numbers of
SFR event counts, respectively: 15,041 for Helios 1, and 7,981 for
Helios 2 throughout their whole mission periods; 1,480 (1,991) for
Voyager 1 (2) in year 1980 only. The SFR properties derived from
each event set are summarized and presented via means of
statistical analysis. Targeted studies using subsets of events are
performed, especially for the purpose of examining the radial
evolution of selected flux rope properties relevant to other
works. Such a study is made feasible by including all derived
event sets distributed over the range of heliocentric distances,
namely, $r\in$ [0.3, 7] au, and the unique approach of
characterizing SFRs by the amount of poloidal magnetic flux
obtained through the GS reconstruction method. The main findings
are summarized as follows.

\begin{enumerate}
\item The event occurrence rate is still on the order of a few
hundreds per month, for the range of radial distances between 0.3
au and 7-8 au near the ecliptic plane.
\item The duration and scale size distributions of SFRs again
exhibit power laws. They possess different power-law slopes at
different radial distances.
\item The axis orientations of the identified cylindrical SFRs
have broad distribution peaks grossly centered around the nominal
Parker spiral field directions at different radial distances. The
trend is more pronounced for the polar angles than for the
azimuthal angles.
\item The bulk properties of SFRs, such as the average magnetic
field magnitude, proton number density and temperature, generally
exhibit clear decay in magnitudes with increasing radial
distances, while the  proton $\langle\beta\rangle$ remains largely
unchanged.
\item The radial changes in magnetic field magnitudes, separately for the  total
field, the axial and the transverse components, seem to follow the
general rule for a Parker magnetic field model, all with a
power-law index close to -1.5.
\item For a uniquely controlled subset of SFR events with the
corresponding unit poloidal magnetic flux $A_m\in$[0.5,1.5]
T$\cdot$m. The radial decaying in $r$ for the following
quantities, event counts $N^2$, average field magnitude $\langle
B\rangle$, and inverse scale size $k$, yields the corresponding
power laws: $N^2\sim r^{-1.5}$, $\langle B\rangle\sim r^{-1.3}$,
and $k\sim r^{-1.1}$.
\end{enumerate}

The radial change in magnetic field seems to be consistent with
the theoretical and observational analysis of 2D MHD turbulence
throughout the heliosphere. For instance, the theoretical work as
well as the analysis of turbulence properties based on in-situ
spacecraft observations by
\citet{Zank2017,Adhikari_2017,Zhao_2017} showed the radial change
of fluctuating magnetic power, following largely a power law $\sim
r^{\gamma}$ with the value $\gamma\approx 2\alpha$ lying between
-2 and -3, consistent with $\alpha\approx$ -1.3 to -1.4 for the
magnetic field variations from our analysis results. In
particular, the radial change of the correlation length scale in
the ranges $r\in[0.3, 1]$ and $>1$ au seems to follow mostly power
laws with different power-law indices
\citep{Adhikari_2017,Zhao_2017}. According to \citet{Zhao_2017},
the correlation length scale changes from about the order of
$\lesssim 0.01$ au for $r\in[0.3,1]$ au to about $\gtrsim 0.01$,
approaching $0.1$ AU for $r\in[1,10]$ au, which is consistent with
the change of the characteristic (average inverse) scale size of
the SFRs from our analysis (see Figure~\ref{fig:r}) over these
radial distances. It is also worth noting that the radial
dependence of $k\sim r^{-1.1}$ should not be interpreted as
indication of self-similar expansion, because this scale
represents the dimension in the radial direction only, not the
lateral (i.e., longitudinal or latitudinal) dimension, which has
the $r^{-1}$ dependence simply due to the radially outward flow.
Therefore such a dependence for the radial dimension may hint at
certain intrinsic processes at work subject to  local conditions
when examined separately at different radial distances. Generally
speaking the apparent effect due to expansion in the radial
dimension is under control by setting the Wal\'en slope threshold
in our approach, with which events with significant remaining
flows, an indication of expansion, are excluded from our event
sets.

Although there appears to be more events from Helios  occurring in
slow solar wind, there is also a trend of increasing Alfv\'enicity
at closer radial distances to the sun, as indicated by the
reduction in event counts (see Table~\ref{tbl:evolution}) for
Helios. A number of newly published studies have indicated the
trend by using the PSP data but with different approaches
\citep[see, e.g.,][]{Kasper2019,2019arXiv191202349Z}.  It is shown
 that the highly Alfv\'enic activity  persisted in rather
low-speed solar wind streams. How the nature of solar wind
fluctuations changes with radial distance or different streams can
be further elucidated  by checking the newly acquired PSP data.
The application of the GS-based automated SFR detection approach
to the publicly available PSP data is currently ongoing.

\acknowledgments We are grateful to our colleagues at SPA/CSPAR,
UAH, Drs.~Laxman Adhikari, Jakobus le Roux,  Gang Li, Gary Webb,
Gary Zank, and Lingling Zhao for on-going collaborations. The
Helios data are accessed via
\url{http://helios-data.ssl.berkeley.edu/}. The other spacecraft
data are provided by the NASA CDAWeb. We acknowledge NASA grants
NNX17AB85G, 80NSSC19K0276, 80NSSC18K0622, and NSF grant
AGS-1650854 for support. Special
 thanks also go to the SCOSTEP/VarSITI program for support of the
 development and maintenance of the on-line small-scale magnetic flux rope
 database website, \url{http://fluxrope.info}.

 \bibliography{bib_database}

\end{document}